\documentstyle[12pt]{article}
\begin{document}

\title{The conservation of the Hamiltonian structures in Whitham's
method of averaging.}

\author{A. Ya. Maltsev.}

\date{L.D.Landau Institute for Theoretical Physics, Kosygina 2,
Moscow 117940, maltsev@itp.ac.ru}

\maketitle

\begin{abstract}
The work is devoted to the proof of the conservation of
local field-theoretical Hamiltonian structures in Whitham's
method of averaging. The consideration is based on the
procedure of averaging of local Poisson bracket, proposed
by B.A.Dubrovin and S.P.Novikov. Using the Dirac procedure
of restriction of the Poisson bracket on the submanifold
in the functional space, it is shown in the generic case
that the Poisson bracket, constructed by method of
Dubrovin and Novikov, satisfies the Jacobi identity.
Besides that, the invariance of this bracket
with respect to the choice of the set of local conservation
laws, used in this procedure, is proved.
\end{abstract}

\begin{center}
{\bf Introduction.}
\end{center}

This work is devoted to Whitham's method of averaging, which permits to
explore the evolution of slow modulated m - phase solutions of
nonlinear systems of equations (~\cite{with}, see also ~\cite{luke},
~\cite{dn2},~\cite{dn3},~\cite{dm}).
The slow modulated parameters of m - phase
solutions (for example "running waves" if $m=1$) satisfy in this approach
to quasilinear homogeneous evolution system of type:
\begin{equation}
\label{zed}
U^{i}_{T} = V^{i}_{j}({\bf U}) U^{j}_{X} , \,\,\,\,\,
i,j = 1,\dots,N ,
\end{equation}
${\bf U} = (U^{1},\dots,U^{N})$.

For the exploring of systems of this kind it appears to be important
to explore them on the subject of being Hamiltonian with respect to
local Poisson brackets of hydrodynamic type
(see~\cite{dn1},~\cite{dn2},~\cite{dn3}).
The theory of these brackets, constructed by B.A.Dubrovin and S.P.Novikov,
(~\cite{dn1},~\cite{dn2},~\cite{dn3}), has been used by S.P.Tsarev
(see ~\cite{tsarev}) for integration of systems (\ref{zed}), having
Hamiltonian structure and reducible to the diagonal form.
The investigation of Hamiltonian
systems which do not satisfy to the last condition,
was made by E.V.Ferapontov
(see ~\cite{fer1},~\cite{fer2}).

In the work ~\cite{dn2} the procedure of constructing of Hamiltonian
structure of the required type for
Whitham's system of equations (\ref{zed}),
under the condition that the initial system is Hamiltonian with respect
to local field-theoretical Poisson brackets was proposed. However,
the proof of the Jacobi identity for constructed by this way brackets
was absent (see ~\cite{novmal}). In this work we shall prove the Jacobi
identity for constructed by Dubrovin-Novikov's method brackets
in the case of "generic situation" with the aid of Dirac's procedure
of restriction of Poisson brackets on the submanifold in the space of
functions.
So that, the results of this work permit to make a statement about the
conservation of local Hamiltonian structures in Whitham's
method of averaging.

\begin{center}
{\bf 1. General constructions.}
\end{center}

Consider the evolution system on the space of fields
$\mbox{$\boldmath \varphi$} = (\varphi^{1},\dots ,\varphi^{n})$ of type:
\begin{equation}
\label{system}
\varphi^{i}_{t} = Q^{i}(\mbox{$\boldmath \varphi$},
\mbox{$\boldmath \varphi$}_{x},\mbox{$\boldmath \varphi$}_{xx},\dots) ,
\,\,\,\,\,i = 1,\dots ,n ,
\end{equation}
which is Hamiltonian with respect to local field-theoretical Poisson
bracket of type:
\begin{equation}
\label{bracket}
\{\varphi^{i}(x) , \varphi^{j}(y)\} = \sum_{k \geq 0} B_{k}^{ij}
(\mbox{$\boldmath \varphi$},\mbox{$\boldmath \varphi$}_{x},\dots )
\delta^{(k)}(x-y)
\end{equation}
(there is a finite number of terms in the sum) with the Hamiltonian:
\begin{equation}
\label{hamilt}
H = \int {\cal P}_{H}(\mbox{$\boldmath \varphi$},
\mbox{$\boldmath \varphi$}_{x},\dots) dx .
\end{equation}

The bracket (\ref{bracket}) can be compatible with the
operator of translation, that is, there
may exist the local functional:
\begin{equation}
\label{momentum}
P = \int{\cal P}_{P}(\mbox{$\boldmath \varphi$},
\mbox{$\boldmath \varphi$}_{x},\dots) dx  ,
\end{equation}
(the momentum operator) such that:
$\{\varphi^{i}(x) , P\} = \varphi^{i}_{x}$ .
(This is not a necessary condition and as was pointed to the
author by O.I.Mokhov, there is a special class of local Poisson
brackets which satisfy it, see ~\cite{mokhov1},~\cite{mokhov2}.
However, in the Hamiltonian structures,
connected with the "physical" systems, this property, as a rule,
is present.)
Besides that, we shall admit that bracket (\ref{bracket}) can have
a finite number of annihilators, that is, functionals (not necessary
having the form (\ref{hamilt}),(\ref{momentum})) $N_{1},\dots,N_{p}$ ,
such that: $\{\varphi^{i}(x) , N_{q}\} = 0$ .

Definition. \newline
Let we are given a function of m variables
$\mbox{$\boldmath \Phi$}(\theta)
= (\Phi^{i}(\theta_{1},\dots,\theta_{m})),
i=1,\dots,n ,  2\pi -$ periodic with respect to each of the arguments,
and corresponding to it m - vectors:
$\mbox{$\boldmath \omega$} = (\omega^{1},\dots,\omega^{m}) ,
{\bf k} = (k^{1},\dots,k^{m})$,
such that:
\begin{equation}
\label{sol}
\omega^{\alpha}\Phi^{i}_{\theta^{\alpha}} =
Q^{i}(\mbox{$\boldmath \Phi$},k^{\alpha}
\mbox{$\boldmath \Phi$}_{\theta^{\alpha}},\dots)
\end{equation}
then the corresponding to them function:
\begin{equation}
\label{resh}
\mbox{$\boldmath \varphi$}(x,t) = \mbox{$\boldmath \Phi$}
({\bf k}x+\mbox{$\boldmath \omega$}t)
\end{equation}
we shall call the m - phase quasiperiodic solution of the system
(\ref{system}).

Note that the existence of m - phase solutions with $m>1$ presents as
a rule only for "integrable" systems like KdV.

System (\ref{sol}) is a system of differential equations in partial
derivatives on functions $\mbox{$\boldmath \Phi$}(\theta)$
$(\theta = (\theta^{1},\dots,\theta^{m}))$,
and its $2\pi -$ periodic with respect to all variables solutions
(if they exist) at all possible ${\bf k}$
and $\mbox{$\boldmath \omega$}$
give the full family of m - phase solutions of system (\ref{system}).

We shall suppose that at any ${\bf k}$ and $\mbox{$\boldmath \omega$}$
in some open set
(we suppose that system (\ref{system}) is not linear) the system
(\ref{system}) has some family of the required solutions, which can be
parametrized by the initial phase shifts $\theta^{\alpha}_{0}$, and
may be also by some set of additional parameters $r^{1},\dots,r^{g}$,
(the number of which is constant at all $\mbox{$\boldmath \omega$}$
and ${\bf k}$), which do not change under the variation of
$\theta^{\alpha}_{0}$. So that, the m - phase solutions of (\ref{system})
will be given by the values
$\mbox{$\boldmath \omega$} = (\omega^{1},\dots,\omega^{m}),
{\bf k} = (k^{1},\dots,k^{m}), r^{1},
\dots,r^{g}$ and $\theta^{1}_{0},\dots,\theta^{m}_{0}$.

Let now the system (\ref{system}) has $N = 2m+g$
translation-invariant functionals:
\begin{equation}
\label{laws}
I^{\nu} = \int {\cal P}^{\nu}(\mbox{$\boldmath \varphi$},
\mbox{$\boldmath \varphi$}_{x},\dots)dx ,
\,\,\,\,\,\nu = 1,\dots,N ,
\end{equation}
commuting with the Hamiltonian (\ref{hamilt}) and being in the
involution with each other with respect to the bracket (\ref{bracket}):
\begin{equation}
\label{invol}
\{I^{\nu} , I^{\mu}\} = 0 .
\end{equation}
Among the integrals (\ref{laws}) there may be Hamiltonian (\ref{hamilt}),
functional of momentum and annihilators of brackets (\ref{bracket})
having the form (\ref{laws}).

We shall assume that parameters
${\bf k}, \mbox{$\boldmath \omega$}$ and ${\bf r} = (r^{1},\dots,r^{g})$
are independent on the family of m - phase solutions of (\ref{system})
and, besides that, may be expressed in terms of the values of integrals
(\ref{laws}) on the functions of family (\ref{resh}), that is
${\bf k} = {\bf k}({\bf U}), \mbox{$\boldmath \omega$} =
\mbox{$\boldmath \omega$}({\bf U}), {\bf r} = {\bf r}({\bf U}) ,
{\bf U} = (U^{1},\dots,U^{N})$, where
\begin{equation}
\label{parameters}
U^{\nu}\! = \lim_{T\rightarrow\infty} {1 \over 2T}\! \int^{T}_{-T}
{\cal P}^{\nu}(\mbox{$\boldmath \varphi$},
\mbox{$\boldmath \varphi$}_{x},\dots)dx =  {1 \over (2\pi)^{m}}
\int_{0}^{2\pi}\!\!\!\!\dots\int_{0}^{2\pi}\!\!
{\cal P}^{\nu}(\mbox{$\boldmath \Phi$},k^{\alpha}\mbox{$\boldmath \Phi$}_
{\theta^{\alpha}},\dots) d^{m}\theta
\end{equation}
(it can be easily seen that values in (\ref{parameters}) do not depend
upon the initial phase shifts $\theta^{\alpha}_{0}$).
So that, after the choice at each value
of parameters ${\bf U}$ on ${\cal M}$
of some definite function $\mbox{$\boldmath \Phi$}_{in}(\theta,{\bf U})$
as having zero initials phases $\theta_{0}$, we can put values $U^{\nu}$
and $\theta^{\alpha}_{0}$ to be the parameters on the family of
m - phase solutions of (\ref{system}).

Each of the functionals (\ref{laws}) generates the Hamiltonian flow of
the form:
\begin{equation}
\label{fluxes}
\varphi^{i}_{\tau^{\nu}} = Q^{i}_{\nu}(\mbox{$\boldmath \varphi$},
\mbox{$\boldmath \varphi$}_{x},\dots) .
\end{equation}

 All these flows are commutative with each other and with (\ref{system}),
and leave the family of m - phase solutions of (\ref{system}) invariant,
generating the linear dependence of the initial phases upon the "times"
$\tau^{\nu}$ on it and leaving constant all the other parameters.
It means that there exist the functions $\omega^{\alpha}_{\nu}({\bf U})$,
such that for all $\mbox{$\boldmath \Phi$}(\theta + \theta_{0},{\bf U})$,
corresponding to the family of m - phase solution (\ref{system}),
we have:
\begin{equation}
\label{fom}
\omega^{\alpha}_{\nu}({\bf U})\Phi^{i}_{\theta^{\alpha}} = Q^{i}_{\nu}
(\mbox{$\boldmath \Phi$},k^{\alpha}\mbox{$\boldmath \Phi$}_
{\theta^{\alpha}},\dots)
\end{equation}

 From (\ref{invol}) we can conclude that all
evolution systems (\ref{fluxes}),
generated by integrals $I^{\nu}$,
have the same properties that the system
(\ref{system}), and all the statements,
proved for system (\ref{system}),
are valid for systems (\ref{fluxes})
(besides zero flows and operator of
shift of $x$, we shall speak about this
in the consideration of Whitham's
method in which the nonlinearity of the system is essential).

System (\ref{sol}) at each $\mbox{$\boldmath \omega$}$ and
${\bf k}$ defines some "submanifold" (let denote it by
${\cal M}_{\mbox{$\boldmath \omega$},{\bf k}}$,
in the space of $2\pi -$ periodic with respect
to each $\theta^{\alpha}$ functions. Functionals
\begin{equation}
\label{sv}
F^{i}_{\mbox{$\boldmath \omega$},{\bf k}}
[\mbox{$\boldmath \Phi$}](\theta)
= \omega^{\alpha}\Phi^{i}_{\theta^{\alpha}}
- Q^{i}(\mbox{$\boldmath \Phi$},k^{\alpha}\mbox{$\boldmath \Phi$}_
{\theta^{\alpha}},\dots)
\end{equation}
are being the constraints, defining these submanifolds.

We shall assume that, if the operator $I^{\nu}$ is not
the operator of momentum or annihilator of (\ref{bracket}), the
corresponding systems (\ref{fom}) at all
$\mbox{$\boldmath \omega$}_{\nu}$ and ${\bf k}$ define (as well as
(\ref{sol})) the full family of m - phase solutions of (\ref{system}),
that is, the constraints:
\begin{equation}
\label{svz}
F^{i (\nu)}_{\mbox{$\boldmath \omega$}_
{\nu},{\bf k}}[\mbox{$\boldmath \Phi$}](\theta) =
\omega^{\alpha}_{\nu}\Phi^{i}_{\theta^{\alpha}} -
Q^{i}_{\nu}(\mbox{$\boldmath \Phi$},
k^{\alpha}\mbox{$\boldmath \Phi$}_{\theta^{\alpha}},\dots)  = 0 ,
\end{equation}
being consider at some definite $\mbox{$\boldmath \omega$}_{\nu}$ and
${\bf k}$, define some $m+g$ - parametric family
${\cal M}^{\nu}_{\mbox{$\boldmath \omega$}_{\nu},{\bf k}}$ of
$2\pi -$ periodic with respect to each $\theta^{\alpha}$ solutions,
parametrized by initial phase shifts and some additional parameters
the number of which is $g$, so that the joint of all these
${\cal M}^{\nu}_{\mbox{$\boldmath \omega$}_{\nu},{\bf k}}$
at all possible $\mbox{$\boldmath \omega$}_{\nu}$ and ${\bf k}$
gives the full family of m - phase solutions of (\ref{system})
(it is so in many known examples).

We shall also assume in our situation (it is also valid in the
examples), that the number g of parameters $r^{1},\dots,r^{g}$ is equal
to the number p of annihilators of (\ref{bracket}), for which there
exist the following motivations:

any of $m+1$ flows (\ref{fluxes}), generated by arbitrary $m+1$
integrals (\ref{laws}), are linearly dependent because of
(\ref{fom}) on the family of m - phase solutions of (\ref{system}),
that is, there exist $\lambda^{1}({\bf U}),\dots,\lambda^{m+1}({\bf U})$,
such that $\sum_{j=1}^{m+1} (\lambda^{j})^{2} = 1$ and
$\sum_{j=1}^{m+1} \lambda^{j}({\bf U})
\omega^{\alpha}_{\nu_{j}}({\bf U}) = 0, \forall \alpha$ .
>From this we can conclude that m - phase solutions of the system
(\ref{system}) can be also defined by the relation:
$$\delta\sum_{j=1}^{m+1} \lambda^{j}I^{\nu_{j}} = \delta\sum_{q=1}^{p}
\mu ^{q}N_{q} ,$$
(where $N^{q}$ - are the annihilators of (\ref{bracket}), since
the functional $\sum_{j=1}^{m+1} \lambda^{j}({\bf U})I^{\nu_{j}}$
generates the zero flow on the corresponding m - phase solutions),
on the functions:
\begin{equation}
\label{func}
\mbox{$\boldmath \varphi$}(x) = \mbox{$\boldmath \Phi$}({\bf k}x) ,
\end{equation}
where $\mbox{$\boldmath \Phi$}(\theta) =
\mbox{$\boldmath \Phi$}(\theta^{1},\dots,\theta^{m})$ is
$2\pi -$ periodic function of $\theta$. Supposing that at any
${\bf k} = (k^{1},\dots,k^{m}), \mu^{1},\dots,\mu^{p}$ and $\lambda^{1},
\dots,\lambda^{m+1}$, satisfying the relation
$\sum_{j=1}^{m+1}(\lambda^{j})^{2} = 1$, the functional
$\sum_{j=1}^{m+1} \lambda^{j}I^{\nu_{j}} - \sum_{q=1}^{p} \mu^{q}N_{q}$
has on the functions (\ref{func}) the only extremum modulo the
initial phase shifts, we have that all m - phase solutions of (\ref{system})
can be characterized by $2m+p$ independent parameters (except the
initial phase shifts), and, so that, we obtain the formulated statement.

As in the finite-dimensional situation we shall require the submanifolds
${\cal M}_{\mbox{$\boldmath \omega$},{\bf k}}$ and
${\cal M}^{(\nu)}_{\mbox{$\boldmath \omega$}_{\nu},{\bf k}}$ (for
$I^{\nu}$ which are not annihilators of (\ref{bracket}) or momentum
operator) to satisfy some property of regularity, analogous to the
maximality of rank of matrix of constraints derivatives in the
finite-dimensional situation. Namely, let us linearize the functional
(\ref{sv}) ${\bf F}_{\mbox{$\boldmath \omega$},{\bf k}}(\theta) =
(F^{1}_{\mbox{$\boldmath \omega$},{\bf k}}(\theta),\dots,F^{n}_
{\mbox{$\boldmath \omega$},{\bf k}}(\theta))$ (respectively any of
functionals (\ref{svz})) on the solution of system (\ref{fom}),
$\mbox{$\boldmath \Phi$}_{in}(\theta + \theta_{0},U)$ (it will be also
a solution of (\ref{svz})), that is introduce the operator
${\hat {\bf L}}_{{\bf U},\theta_{0}}$
(respectively ${\hat {\bf L}}^{(\nu)}_{{\bf U},\theta_{0}}$)
with the kernel $L^{ij}_{{\bf U}}(\theta + \theta_{0},\theta^{\prime} +
\theta_{0})
\,\,\,(L^{ij(\nu)}_{{\bf U}}(\theta + \theta_{0},
\theta^{\prime} + \theta_{0}))$, such that:
$$\delta F^{i}_{\mbox{$\boldmath \omega$}({\bf U}),{\bf k}({\bf U})}
(\theta) =$$
$$= {1 \over (2\pi)^{m}}
\int_{0}^{2\pi}\!\!\dots\int_{0}^{2\pi} \sum_{j} L^{ij}_{{\bf U}}
(\theta + \theta_{0},\theta^{\prime} + \theta_{0})
(\Phi^{j}(\theta^{\prime}) - \Phi_{in}^{j}(\theta^{\prime} + \theta_{0},
{\bf U}))d^{m}\theta  =$$
\begin{equation}
\label{lin}
= ({\hat {\bf L}}_{{\bf U},\theta_{0}} \delta \mbox{$\boldmath \Phi$})^
{i}(\theta)
\end{equation}
(similar for the constraints $F^{i(\nu)}_{\mbox{$\boldmath \omega$}_
{\nu},{\bf k}}[\mbox{$\boldmath \Phi$}](\theta)$).

Defined by such a way ${\hat {\bf L}}_{{\bf U},\theta_{0}}
\,\,\,({\hat {\bf L}}^{(\nu)}_{{\bf U},\theta_{0}})$
are differential with respect to $\theta$ operators with the periodic
coefficients in the space of $2\pi -$ periodic functions
$(\Phi^{i}(\theta), i=1,\dots,n)$.
We shall require the following conditions:

A) For all ${\bf U}$ and $\theta_{0}$
the kernel of operator (\ref{lin})
${\bf L}_{{\bf U},\theta_{0}}$
consists of vectors tangential to the submanifold
${\cal M}_{\mbox{$\boldmath \omega$}({\bf U}),{\bf k}({\bf U})}$ ,
defined by system ({\ref{fom}),
that is, the functions
$\mbox{$\boldmath \Phi$}_{\theta^{\alpha}}(\theta,
{\bf k},\mbox{$\boldmath \omega$},{\bf r})$ and
$\mbox{$\boldmath \Phi$}_{r^{\zeta}}
(\theta,{\bf k},\mbox{$\boldmath \omega$},{\bf r})$ give the basis
in the space of solutions of system
$$({\hat {\bf L}}_{{\bf U},\theta_{0}}
\delta\mbox{$\boldmath \Phi$})(\theta) = 0 ,$$
and the same property is valid for operators
${\hat {\bf L}}^{(\nu)}_{{\bf U},\theta_{0}}$,
corresponding to $I^{\nu}$, which are not annihilators of (\ref{bracket})
and operator of momentum.

B) The co-dimension of the images of the
operators ${\hat {\bf L}}_{{\bf U},\theta_{0}} ,
{\hat {\bf L}}^{(\nu)}_{{\bf U},\theta_{0}}$
in the space of $2\pi -$ periodic with respect to $\theta$ functions
is equal to the dimension of their kernels, that is, all
${\hat {\bf L}}_{{\bf U},\theta_{0}} ,
{\hat {\bf L}}^{(\nu)}_{{\bf U},\theta_{0}}$ have exactly $m+g\,\, (g=p)$
left eigen vectors (let denote them $\mbox{$\boldmath \kappa$}^{(s)}_
{{\bf U},\theta_{0}}$ and $\mbox{$\boldmath \kappa$}^{(s),(\nu)}_
{{\bf U},\theta_{0}}$), that is $2\pi -$ periodic with respect to each
$\theta^{\alpha}$ functions $\kappa^{(s)}_{i {\bf U}}(\theta + \theta_{0}),
j=1,\dots,n, s=1,\dots,m+g$ and
$\kappa^{(s),(\nu)}_{i {\bf U}}(\theta + \theta_{0})$, such that:
\begin{equation}
\label{lev}
{1 \over (2\pi)^{m}}\int_{0}^{2\pi}\!\!\dots\int_{0}^{2\pi}
\kappa^{(s)}_{i {\bf U}}(\theta) L^{ij}_{{\bf U}}(\theta,\theta^{\prime})
d^{m}\theta \equiv 0
\end{equation}
\begin{equation}
\label{levv}
{1 \over (2\pi)^{m}}\int_{0}^{2\pi}\!\!\dots\int_{0}^{2\pi}
\kappa^{(s),(\nu)}_{i {\bf U}}(\theta) L^{ij(\nu)}_{{\bf U}}(\theta,
\theta^{\prime}) d^{m}\theta \equiv 0 .
\end{equation}

If all these conditions are satisfied, we say that submanifolds
${\cal M}_{\mbox{$\boldmath \omega$},{\bf k}}$ and
${\cal M}^{\nu}_{\mbox{$\boldmath \omega$}_{\nu},k}$ in the space of
$2\pi -$ periodic functions of $\theta$ have the property of regularity.
Besides that, under the assumption that vectors
$\mbox{$\boldmath \Phi$}_{\theta^{\alpha}}(\theta,
{\bf k},\mbox{$\boldmath \omega$},{\bf r}) ,\newline
\mbox{$\boldmath \Phi$}_{r^{\zeta}}(\theta,{\bf k},
\mbox{$\boldmath \omega$},{\bf r}) , \mbox{$\boldmath \Phi$}_{k^{\alpha}}
(\theta,{\bf k},\mbox{$\boldmath \omega$},{\bf r})$ and
$\mbox{$\boldmath \Phi$}_{\omega^{\beta}}(\theta,
{\bf k},\mbox{$\boldmath \omega$},{\bf r})$ are linearly independent
at all values of parameters
$(\theta_{0},{\bf k},\mbox{$\boldmath \omega$},{\bf r})$
(it is essential requirement for the following consideration),
the joint of described above submanifolds
${\cal M}_{\mbox{$\boldmath \omega$},{\bf k}}$
(or ${\cal M}^{\nu}_{\mbox{$\boldmath \omega$}_{\nu},{\bf k}}$)
gives $3m+g = N+m -$ dimensional submanifold ${\cal M}$ in the space of
$2\pi$ - periodic with respect to each $\theta^{\alpha}$ functions,
corresponding to the full family of $m$ - phase solutions
of (\ref{system}).

\begin{center}
{\bf 2. Method of Whitham.}
\end{center}

The constructions described above are closely connected with Whitham's
method of averaging for nonlinear systems of differential equations
in partial derivatives (it can not be applied to the flows, generated
by annihilators of (\ref{bracket}) or momentum operator), which is the
following procedure: introduced the small parameter
$\epsilon$, we put $T = \epsilon t , X = \epsilon x$
and rewrite the system
(\ref{system}) on the fields $\varphi^{i}(X,T)$ in the form:
\begin{equation}
\label{s}
\epsilon \varphi^{i}_{T} = Q^{i} (\mbox{$\boldmath \varphi$},
\epsilon\mbox{$\boldmath \varphi$}_{X},
\epsilon^{2}\mbox{$\boldmath \varphi$}_{XX},\dots) .
\end{equation}

Let now consider the system $(\ref{s})$ on the space of functions
$\mbox{$\boldmath \varphi$}(\theta,X,T) ,
\linebreak \theta = (\theta^{1},\dots,\theta^{m})$ ,
$2\pi$ - periodic with respect to each of variables $\theta^{\alpha}$.
In method of Whitham we try to find the functions:
$${\bf S}(X,T) = (S^{1}(X,T),\dots,S^{m}(X,T)) ,$$
and $2\pi$-periodic
with respect to all $\theta^{\alpha}$ functions:
$$\mbox{$\boldmath \Phi$}(\theta,X,T,\epsilon) =
(\Phi^{i}(\theta,X,T,\epsilon)) ,$$
represented by asymptotical series when $\epsilon \rightarrow 0$:
\begin{equation}
\label{r}
\Phi^{i}(\theta,X,T,\epsilon) =
\sum_{n=0}^{\infty} \epsilon^{n} \Phi^{i}_{(n)}
(\theta,X,T) ,
\end{equation}
such that the function:
\begin{equation}
\label{re}
\mbox{$\boldmath \varphi$} (\theta,X,T,\epsilon) =
\mbox{$\boldmath \Phi$} (\theta + {{\bf S}(X,T) \over
\epsilon} ,X,T,\epsilon) =\sum_{n=0}^{\infty} \epsilon^{n}
\mbox{$\boldmath \Phi$}_{(n)}(\theta +
{{\bf S}(X,T) \over \epsilon} ,X,T,)
\end{equation}
- satisfies the system (\ref{s}) at all $\theta$ and
$\epsilon$, ($\epsilon \rightarrow 0$).

It can be easily seen that the substitution of (\ref{re}) into the
system (\ref{s}) gives the following equations in the zero order
of $\epsilon$
\begin{equation}
\label{ur}
S^{\alpha}_{T} \Phi^{i}_{(0) \theta^{\alpha}}(\theta,X,T)
= Q^{i}(\mbox{$\boldmath \Phi$}_{(0)},
S^{\alpha}_{X}\mbox{$\boldmath \Phi$}_{(0) \theta^{\alpha}},\dots) ,
\end{equation}
that is, at any $X$ and $T$,
$\mbox{$\boldmath \Phi$}_{(0)}(\theta,X,T)$,
as a function of $\theta$, represents one of
the function from family ${\cal M}$ described above,
and the functions $\mbox{$\boldmath \varphi$}(x,t,\epsilon) =
\mbox{$\boldmath \Phi$}(\theta_{0} +
{1 \over \epsilon}{\bf S}(\epsilon x,\epsilon t),\epsilon)$,
obtained from (\ref{re}) after the
replacement of $X$ and $T$ by $\epsilon x$ and $\epsilon t$
respectively, tend at small $\epsilon$ to the slow modulated
m - phase solutions of (\ref{system}).

Besides that, from (\ref{ur}) we obtain:
\begin{equation}
\label{soot}
{\bf k}({\bf U}) = {\bf S}_{X} , \,\,\,\,\,
\mbox{$\boldmath \omega$}({\bf U}) = {\bf S}_{T} ,
\end{equation}
where ${\bf U}, {\bf k}$ and $\mbox{$\boldmath \omega$}$ -
are parameters on ${\cal M}$.

Terms with $\epsilon^{k}, k>0,$ give the relations:
\begin{equation}
\label{n}
({\hat {\bf L}}_{{\bf U(X,T)},\theta_{0}(X,T)}
\mbox{$\boldmath \Phi$}_{(k)})^{i}(\theta,X,T)
= f^{i}_{k}
(\mbox{$\boldmath \Phi$}_{(0)},\mbox{$\boldmath \Phi$}_{(1)},\dots,
\mbox{$\boldmath \Phi$}_{(k-1)},{\bf S}_{X},{\bf S}_{T},\dots)
\end{equation}
- where ${\hat {\bf L}}_{{\bf U},\theta_{0}}$ - is
described in (\ref{lin}) linear differential
(with respect to $\theta$)
operator with $2\pi -$ periodic with respect to
$\theta$ coefficients (expressed in terms of the
$\mbox{$\boldmath \Phi$}_{(0)}(\theta,
\theta_{(0)}(X,T),{\bf U}(X,T))$ and its derivatives with respect to
$\theta^{\alpha}$), ${\bf f}_{k}$ is discrepancy,
which depends upon the previous functions
$\mbox{$\boldmath \Phi$}_{(j)}(\theta,X,T)$ and being of order of
$k$, regarding that functions $\mbox{$\boldmath \Phi$}_{(j)}$
have order $j$, functions ${\bf S}$ - order $-1$, multiplication of
functions adds their orders and differentiation with
respect to $T$ and $X$ adds $1$ to the order of function.

The system (\ref{n}) has solutions in the class of $2\pi$ - periodic
with respect to $\theta$ functions if and only if at any $X$ and $T$
the vector ${\bf f}_{k}(\theta,X,T)$ is orthogonal for all defined in
(\ref{lev}) left eigen vectors of operator
${\hat {\bf L}}_{{\bf U}(X,T),\theta_{0}(X,T)}$,
that is:
\begin{equation}
\label{ort}
{1 \over (2\pi)^{m}}\int_{0}^{2\pi}\!\!\dots\int_{0}^{2\pi}
\kappa^{(s)}_{i {\bf U}(X,T)}(\theta + \theta_{0})
f^{i}_{k}(\theta,X,T)
d^{m}\theta \equiv 0 ,
\end{equation}
which gives on the function ${\bf f}_{k}(\theta,X,T)$
at any $X$ and $T$ $m+g = N-m$ independent relations according to the
number of left eigen vectors of the operator
${\hat {\bf L}}_{{\bf U}(X,T),\theta_{0}(X,T)}$.

Taking $\mbox{$\boldmath \Phi$}_{(0)}$ in the form:
$\mbox{$\boldmath \Phi$}_{(0)}(\theta,X,T) =
\mbox{$\boldmath \Phi$}_{in}(\theta +
\theta_{0}(X,T),{\bf U}(X,T))$,
we can easily see from the statements formulated above that in the
first order of $\epsilon$ equations
$(\ref{n})$ have the form:
$$({\hat {\bf L}}_{[{\bf U}(X,T),\theta_{0}(X,T)]}
\mbox{$\boldmath \Phi$}_{1})^{i}(\theta) =
\sum_{{\bf n}}\alpha^{i}_{j({\bf n})}({\bf S}_{X},\mbox{$\boldmath \Phi$}_
{(0)},\mbox{$\boldmath \Phi$}_{(0)\theta^{\alpha}},
\mbox{$\boldmath \Phi$}_{(0)\theta^{\alpha}\theta^{\beta}},\dots)
\times$$
$$\times\left(\mbox{$\boldmath \Phi$}^{j}_{(0){\bf n}
\theta, U^{\nu}}U^{\nu}_{X} +
\mbox{$\boldmath \Phi$}^{j}_{(0){\bf n}\theta,
\theta^{\alpha}}\theta^{\alpha}_{0X}\right) +
\beta^{i}_{\alpha}({\bf S}_{X},\mbox{$\boldmath \Phi$}_{(0)},
\mbox{$\boldmath \Phi$}_{(0)\theta^{\alpha}},
\mbox{$\boldmath \Phi$}_{(0)\theta^{\alpha}\theta^{\beta}},\dots)
{\bf S}_{XX} -$$
\begin{equation}
\label{dd}
- \left( \mbox{$\boldmath \Phi$}^{i}_{(0)U^{\nu}}U^{\nu}_{T} +
\mbox{$\boldmath \Phi$}^{i}_{(0)\theta^{\alpha}}
\theta^{\alpha}_{0T} \right) ,
\end{equation}
where ${\bf n} = (n_{1},\dots,n_{m})$ is an integral m - vector
with nonnegative components, ${\bf n}\theta$ denotes here
$(n_{1}\theta^{1} \dots n_{m}\theta^{m})$,
$\alpha^{i}_{j{\bf n}}$ и $\beta^{i}_{\alpha}$ - some definite functions,
values ${\bf S}_{X}$ are ${\bf S}_{T}$ are connected with the parameters
${\bf U}(X)$, corresponding to $\mbox{$\boldmath \Phi$}_{(0)}
(\theta,X)$, by the relations (\ref{soot}).
So that the condition (\ref{ort}) gives on the functions ${\bf U}(X,T)$
and $\theta_{0}(X,T)$ $N-m$ equations of the form:
\begin{equation}
\label{sss}
A^{\zeta}_{\mu}({\bf U}) U^{\mu}_{T} + B^{\zeta}_{\mu}({\bf U})
U^{\mu}_{X} + A^{\prime \zeta}_{\alpha}({\bf U}) \theta_{0T}^{\alpha}
+ B^{\prime \zeta}_{\alpha}({\bf U}) \theta_{0X}^{\alpha} = 0, \,\,\,
\zeta = 1,\dots,N-m .
\end{equation}
Adding to them m equations:
\begin{equation}
\label{scv}
k^{\alpha}_{T} = \omega^{\alpha}_{X} ,
\end{equation}
following from (\ref{soot}), we obtain the system of $N$ quasilinear
equations on $N+m$ functions $U^{\nu}(X,T) , \theta_{0}^{\alpha}(X,T)$ .

If the conditions (\ref{sss}) and (\ref{scv})
are satisfied, the function
$\mbox{$\boldmath \Phi$}_{(1)}(\theta,X,T)$
can be found at any $X$ and $T$
from the differential with respect to $\theta$ equation modulo the
arbitrary linear combination of $N-m$ functions described in (A)
and lying in the
kernel of the operator ${\hat {\bf L}}_{{\bf U}(X,T),\theta_{0}(X,T)}$
(let here denote them as
$\mbox{$\boldmath \xi$}^{(s)}_{{\bf U},\theta_{0}}$),
that is, modulo the function of type:
$\sum_{s=1}^{N-m} C_{s}(X,T) \mbox{$\boldmath \xi$}^{(s)}_
{{\bf U}(X),\theta_{0}(X)}$.

The values of the coefficients $C_{s}(X,T)$ 
can be found from the condition
of the solvability of system (\ref{n}) in the next order of
$\epsilon$, where we have the same situation and, so that,
if (\ref{sss}) and (\ref{scv}) are satisfied, we can obtain
sequentially all the terms of series (\ref{r}), which permits to find
the required functions $\mbox{$\boldmath \Phi$}(\theta,X,T,\epsilon)$.
Function ${\bf S}(X,T)$ can be constructed using the values
${\bf k}({\bf U})$ and $\mbox{$\boldmath \omega$}({\bf U})$.

   Lemma 1.

Under the conditions formulated above, that is, in the presence of $N$
local translation-invariant integrals (\ref{laws}), such that the
evolution of their densities ${\cal P}^{\nu}(\mbox{$\boldmath \varphi$},
\mbox{$\boldmath \varphi$}_{x},\dots)$ according
to the flow (\ref{system}) has the form:
\begin{equation}
\label{pot}
{\cal P}^{\nu}_{t} = {\cal R}^{\nu}_{x}(\mbox{$\boldmath \varphi$},
\mbox{$\boldmath \varphi$}_{x},\dots)
\end{equation}
with some ${\cal R}^{\nu}$, the system (\ref{sss}) does not contain
$\theta_{0}(X,T)$ and provides the relations only for ${\bf U}(X,T)$,
that is, all the terms containing $\theta_{0X}$ and $\theta_{0T}$
are orthogonal to the left eigen
vectors of the operator ${\hat {\bf L}}$
independently upon the values of parameters ${\bf U}(X,T)$ and
$\theta_{0}(X,T)$, so that
$A^{\prime \zeta}_{\mu} \equiv 0 , B^{\prime \zeta}_{\mu} \equiv 0$.
Besides that, in method of Whitham for any of the systems (\ref{fluxes}),
which is not the operator of translation or trivial flow, we have the
same situation.

Proof.

If (\ref{sss}) and (\ref{scv}) are valid, there exist the solutions of
(\ref{s}) in form of asymptotic series (\ref{r}),
the substitution of which
into (\ref{pot}) and integration with respect to $\theta$ gives in
the first order of $\epsilon$:
\begin{equation}
\label{up}
U^{\nu}_{T} = \partial_{X} \langle {\cal R}^{\nu} \rangle({\bf U}) ,
\end{equation}
- where $\langle \dots \rangle$ means the averaging on the family
${\cal M}$ defined by the formula:
\begin{equation}
\label{us}
\langle F(\mbox{$\boldmath \varphi$},
\mbox{$\boldmath \varphi$}_{x},\dots) \rangle ({\bf U}) \equiv
{1 \over (2\pi)^{m}}\int_{0}^{2\pi}\!\!\dots\int_{0}^{2\pi}
F(\mbox{$\boldmath \Phi$},k^{\alpha}\mbox{$\boldmath \Phi$}_
{\theta^{\alpha}},k^{\alpha}k^{\beta}\mbox{$\boldmath \Phi$}_
{\theta^{\alpha}\theta^{\beta}},\dots) d^{m}\theta ,
\end{equation}
if $\mbox{$\boldmath \varphi$}(x) = \mbox{$\boldmath \Phi$}_
{in}({\bf k} x + \theta_{0},{\bf U})$.

So that, from the system (\ref{sss})-(\ref{scv})
follows the system (\ref{up}), which is a system of $N$ independent
in the general case equations on functions ${\bf U}(X,T)$,
having the same form as (\ref{sss})-(\ref{scv}),
and, so that, (\ref{up}) is equivalent to (\ref {sss})-(\ref{scv}).
Since these conclusions may be applied to any of systems
(\ref{fluxes}), which is not translation or trivial flow,
the Lemma is proved.

Let us note that in the presence of the additional
conservation laws of form (\ref{pot}) their averaging gives
the equations which are the corollaries of (\ref{up}),
the independence of (\ref{up}), which is the
system of $N$ equations on $N$ parameters $U^{\nu}(X)$,
is the property of generic situation.

System (\ref{up}) (or equivalent to it (\ref{sss})-(\ref{scv})),
giving the evolution of slow modulated parameters ${\bf U}(X)$
of m - phase solutions of (\ref{system}) is called the system
of equations of Whitham.

System (\ref{up}) has the form:
\begin{equation}
\label{uu}
U^{\nu}_{T} = V^{\nu}_{\mu}({\bf U}) U^{\mu}_{X} , \,\,\,
\nu,\mu  =1,\dots,N ,
\end{equation}
that is, it refers to evolution systems of hydrodynamic type.
For smooth initial dates ${\bf U}(X)$ system (\ref{uu})
has in general case smooth solution up to some moment $T_{0}$
depending upon the initial dates and after that the solution
will be broken. So that we can use the solutions of Whitham
up to the definite point of time, after which
they are not defined.

Let us point also that usually we consider just
the finite number of terms of asymptotic series (\ref{r})
when they give us solutions of initial system (\ref{system})
modulo terms of higher order of $\epsilon$. In particular
it is possible to consider just the first term of (\ref{r}),
which gives the evolution of slow-modulated m-phase solutions
of (\ref{system}) provided that the Whitham equations on
parameters ${\bf U}$ hold and the next term in (\ref{r})
is uniformly bounded. We shall consider all arising here
asymptotic series from this point of view, in particular,
we shall not interest in there convergency regions since
we consider the relations on the first (say k) terms of such series
modulo the higher orders of $\epsilon$.\footnote{Author
is grateful to I.M.Krichever for fruitful discussions of
these questions.}

\begin{center}
{\bf 3. The Poisson brackets of the hydrodynamic type.}
\end{center}

Among the systems (\ref{uu}) there is an especial class of them
which are Hamiltonian with respect to Poisson
brackets of type:
\begin{equation}
\label{br}
\{U^{\nu}(X) , U^{\mu}(Y)\} = 
g^{\nu\mu}({\bf U}(X)) \delta^{\prime}(X-Y) +
b_{\lambda}^{\nu\mu}({\bf U}(X))U^{\lambda}_{X} \delta(X-Y)
\end{equation}
with local Hamiltonian of hydrodynamic type:
\begin{equation}
\label{gg}
H = \int h({\bf U}(X))dx ,
\end{equation}
which plays an important role in their integrability
(see ~\cite{tsarev},~\cite{fer1},~\cite{fer2}).

The theory of brackets (\ref{br}) with nondegenerated 
$g^{\nu\mu}({\bf U})$,
constructed by B.A.Dubrovin and S.P. Novikov
(see ~\cite{dn1},~\cite{dn2},~\cite{dn3}),
is closely connected with Riemannian geometry. In particular, from their
skew-symmetry follows:
\begin{equation}
\label{rg}
g^{\nu\mu} = g^{\mu\nu} , \,\,\,\,\, b_{\lambda}^{\nu\mu} + b_{\lambda}^
{\mu\nu} = {\partial g^{\nu\mu} \over \partial U^{\lambda}} ,
\end{equation}
Leibnitz identity leads to the fact that under the transformations
of coordinates $U^{\nu} \rightarrow {\tilde U}^{\nu}({\bf U})$ functions
$g^{\nu\mu}$ transform as the contravariant components of a metric
tensor, whereas the functions
$\Gamma^{\nu}_{\mu\lambda} = - g_{\mu\tau} b_{\lambda}^{\tau\nu}$
($g_{\mu\tau}g^{\tau\nu} = \delta^{\nu}_{\mu}$) transform as coefficients
of connection, consistent with metric view (\ref{rg}).
The Jacobi identity for (\ref{br}) in the case of nondegenerated metric
$g^{\nu\mu}$ is equivalent to the symmetry of connection
$\Gamma^{\nu}_{\mu\lambda}$ and zero curvature of the metric:
$R^{\nu}_{\mu\lambda\tau} \equiv 0$.

The theory of brackets (\ref{br}) with degenerated 
$g^{\nu\mu}({\bf U})$ is more complicated, but also has
a nice geometrical form, see ~\cite{grinberg}.

In ~\cite{dn2} B.A.Dubrovin and S.P.Novikov proposed also the
method of constructing of the brackets (\ref{br}) for Whitham's system of
equations (\ref{uu}), starting from the Hamiltonian structure
(\ref{bracket}) for the initial system (\ref{system}) under the
condition that we have the necessary number of commuting integrals
(\ref{laws}). Namely, let us calculate the brackets of densities of
integrals (\ref{laws}) in the form:
\begin{equation}
\label{skob}
\{{\cal P}^{\nu}(\mbox{$\boldmath \varphi$}(x),
\mbox{$\boldmath \varphi$}_{x},\dots),{\cal P}^{\mu}
(\mbox{$\boldmath \varphi$}(y),\mbox{$\boldmath \varphi$}_{y},\dots)\} =
\sum_{k\geq0} A_{k}^{\nu\mu}(\mbox{$\boldmath \varphi$}(x),
\mbox{$\boldmath \varphi$}_{x},\dots) \delta^{k}(x-y) ,
\end{equation}
(there is a finite number of terms in the sum, $\nu,\mu = 1,\dots,N$).

View (\ref{invol}) we can conclude that
\begin{equation}
\label{aq}
A_{0}^{\nu\mu}(\mbox{$\boldmath \varphi$},
\mbox{$\boldmath \varphi$}_{x},\dots) \equiv \partial_{x} Q^{\nu\mu}
(\mbox{$\boldmath \varphi$},\mbox{$\boldmath \varphi$}_{x},\dots)
\end{equation}
for some $Q^{\nu\mu}$. In the described above coordinates
${\bf U} = (U^{1},\dots,U^{N})$ the brackets of Dubrovin-Novikov have the
form:
\begin{equation}
\label{sdn}
\{U^{\nu}(X) , U^{\mu}(Y)\} = \langle A_{1}^{\nu\mu} \rangle ({\bf U}(X))
\delta^{\prime}(X-Y) + {\partial \langle Q^{\nu\mu} \rangle ({\bf U}(X))
\over \partial X} \delta(X-Y)
\end{equation}
-where $\langle\dots\rangle$, as previously, denotes the
averaging on the m - phase solutions of (\ref{system}),
defined by formula (\ref{us}).

However, the proof of the fact that constructed by such
a way brackets satisfy the Jacobi identity, was absent
in ~\cite{dn2} (see ~\cite{novmal}).
The main purpose of this paper is to prove the fact that the procedure
(\ref{skob}) - (\ref{sdn}) of averaging of brackets (\ref{bracket})
really gives the Poisson brackets of type (\ref{br}), satisfying the
Jacobi identity.

We shall need for the further purposes the Dirac procedure of
restriction of Poisson bracket on the submanifold. Let us describe
here this procedure in the notations of spaces of finite dimension.

Let in the space $V$ with the coordinates
${\bf x} = (x^{1},\dots,x^{l})$ and the Poisson bracket:
\begin{equation}
\label{skp}
\{x^{g},x^{p}\} = J^{qp}({\bf x}) ,
\end{equation}
be a submanifold $N$, defined with the aid of
constraints $g^{1},\dots,g^{s}$ by the equations:
\begin{equation}
\label{pdmn}
g^{1}({\bf x}) = 0 , \,\,\dots\,\, , g^{s}({\bf x}) = 0 .
\end{equation}
Let us suppose also that some $l-s$ functions
$f^{1}({\bf x}),\dots,f^{l-s}({\bf x})$, defined in the vicinity of
$N$ in the space $V$, give the coordinate system on $N$ after the
restriction on it. All the redefinitions of the functions
$f^{\lambda}({\bf x})$, having the form
$${\tilde f}^{\lambda}({\bf x}) =
f^{\lambda}({\bf x}) + \sum_{\zeta=1}^
{s} \tau^{\lambda}_{\zeta}({\bf x}) g^{\zeta}({\bf x})$$
with arbitrary $\tau^{\lambda}_{\zeta}({\bf x})$, do not change
this coordinate system. Let us suppose that we can find functions
$\tau^{\lambda}_{\zeta}({\bf x})$, such that the submanifold $N$ is
invariant under Hamiltonian flows
generated by all the functions ${\tilde f}^{\lambda}({\bf x})$
(in the case of nondegeneracy of the matrix
$\{g^{\zeta}({\bf x}),g^{\xi}({\bf x})\}$ on $N$ it is always possible),
that is:
$\{{\tilde f}^{\lambda}({\bf x}),g^{\zeta}({\bf x})\} \equiv 0$ if
${\bf g}({\bf x}) = 0$. Then we can define the Dirac restriction
of bracket (\ref{skp}) on the submanifold (\ref{pdmn}) by the formula:
\begin{equation}
\label{opd}
\{f^{\lambda},f^{\mu}\}^{*} = \{{\tilde f}^{\lambda}({\bf x}) ,
{\tilde f}^{\mu}({\bf x})\}|_{N}({\bf f})
\end{equation}
The bracket (\ref{opd}) satisfies automatically all the necessary
identities. It can be easily verified that with the functions
$\tau^{\lambda}_{\zeta}({\bf x})$, founded by such a way, the bracket
(\ref{opd}) may be also written in the form:
\begin{equation}
\label{ogran}
\{f^{\lambda},f^{\mu}\}^{*} =
\{f^{\lambda}({\bf x}),f^{\mu}({\bf x})\}|_{N}({\bf f})
- \left(\tau^{\lambda}_{\zeta}({\bf x})\tau^{\mu}_{\xi}({\bf x})
\{g^{\zeta}({\bf x}),g^{\xi}({\bf x})\}\right)|_{N}({\bf f}) .
\end{equation}
The infinite dimensional form of (\ref{ogran}) will be very convenient
in the further analysis.

\begin{center}
{\bf 4. The coordinates in the vicinity of submanifold
corresponding to the full family of m-phase solutions.}
\end{center}

Let us now return to the constructions connected with Whitham's method
of averaging. After the substitution $X = \epsilon x$ the bracket
(\ref{bracket}) has the form:
\begin{equation}
\label{ebr}
\{\varphi^{i}(X),\varphi^{j}(Y)\} = \sum_{k\geq 0}
B^{ij}_{k}(\mbox{$\boldmath \varphi$},
\epsilon\mbox{$\boldmath \varphi$}_{XX},
\epsilon^{2}\mbox{$\boldmath \varphi$}_{XX},\dots)
\epsilon^{k}\delta^{(k)}(X-Y) ,
\end{equation}
integrals (\ref{laws}) become:
\begin{equation}
\label{elaws}
I^{\nu} = \epsilon^{-1}\int {\cal P} (\mbox{$\boldmath \varphi$},
\epsilon\mbox{$\boldmath\varphi$}_{X},\dots) dX
\end{equation}
Let us consider the space of functions
$\mbox{$\boldmath \varphi$}(\theta^{1},\dots,\theta^{m},X)$,
$2\pi -$ periodic with respect to each of $\theta^{\alpha}$ and define
at all $\epsilon$ the Poisson bracket on it according to the formula:
\begin{equation}
\label{esk}
\{\varphi^{i}(\theta,X),
\varphi^{j}(\theta^{\prime},Y)\} =
\sum_{k\geq 0}B^{ij}_{k} (\mbox{$\boldmath \varphi$},
\epsilon\mbox{$\boldmath \varphi$}_{X},
\epsilon^{2}\mbox{$\boldmath \varphi$}_{XX},\dots)
\epsilon^{k} \delta^{(k)}(X-Y) \delta(\theta - \theta^{\prime})
\end{equation}
(where $\delta(\theta-\theta^{\prime})$ is the $\delta -$ function
in $m -$ dimensional space).

Consider in the space of $2\pi -$ periodic with respect to $\theta$
functions the submanifold ${\cal M}^{\prime}$ of functions
$\mbox{$\boldmath \varphi$}(\theta,X)$, such that at any
$X$ $\mbox{$\boldmath \varphi$}(\theta,X)$
as a function of $\theta$ lies in ${\cal M}$.
The functions $U^{\nu}(X)$ and $\theta_{0}^{\alpha}(X)$
can be taken as the coordinates on the submanifold
${\cal M}^{\prime}$, so that the functions
$\mbox{$\boldmath \varphi$}(\theta,X)$
from ${\cal M}^{\prime}$ will be represented by the formula:
\begin{equation}
\label{fms}
\mbox{$\boldmath \varphi$} (\theta,X) =
\mbox{$\boldmath \Phi$}_{in}(\theta +
\theta_{0}(X), {\bf U}(X))  .
\end{equation}

After the prolongation of coordinates ${\bf U}(X)$ and $\theta_{0}(X)$
by the independent upon $\epsilon$ way
in the vicinity of ${\cal M}^{\prime}$
(let us denote it by $\Delta_{\delta}$) in the functional space of
$2\pi -$ periodic with respect to $\theta$ functions,
where these functions satisfy the conditions:
$$\mbox{$\boldmath \varphi$}(\theta,X) \in
\Delta_{\delta} \Leftrightarrow  \exists
\mbox{$\boldmath \varphi$}_{0}(\theta,X)
\in {\cal M}^{\prime} :
\|\mbox{$\boldmath \varphi$}(\theta,X) -
\mbox{$\boldmath \varphi$}_{0}(\theta,X)\| \equiv$$
$$\equiv \sum_{i}( max |\varphi^{i}(\theta,X) -
\varphi_{0}^{i}(\theta,X)| + {1 \over 1!}
(\sum_{\alpha}max |\varphi^{i}_{\theta^{\alpha}} -
\varphi^{i}_{0 \theta^{\alpha}}| +
max |\varphi^{i}_{X} - \varphi^{i}_{0X}|) +$$
\begin{equation}
\label{del}
+{1 \over 2!}(\sum_{\alpha,\beta} max |\varphi^{i}_
{\theta^{\alpha}\theta^{\beta}} - \varphi^{i}_{0 \theta^{\alpha},
\theta^{\beta}}| + \sum_{\alpha} max |\varphi^{i}_{\theta^{\alpha}X} -
\varphi^{i}_{0 \theta^{\alpha}X}| +
max |\varphi^{i}_{XX} - \varphi^{i}_{0 XX}|) +
\end{equation}
$$+ {1 \over 3!}(\dots) + \dots) < \delta$$
(that is we imply that there exists a function
$\mbox{$\boldmath \varphi$}_{0}(\theta,X)$ from
${\cal M}^{\prime}$ such that all the terms of the series (\ref{del})
exist, the series (\ref{del}) is convergent
and satisfies to the formulated
condition), we can define the submanifold ${\cal M}^{\prime}$ with the
aid of constraints:
\begin{equation}
\label{fsv}
F^{i}(\theta,X)[\mbox{$\boldmath \Phi$}] =
\omega^{\alpha}({\bf U}[\mbox{$\boldmath \varphi$}](X))\varphi^{i}_
{\theta^{\alpha}} - Q^{i}(\mbox{$\boldmath \varphi$},
k^{\alpha}({\bf U}[\mbox{$\boldmath \varphi$}](X))\varphi_
{\theta^{\alpha}},\dots)= 0 .
\end{equation}
The difference between the systems (\ref{sol}) and (\ref{fsv}) is that the
system (\ref{fsv}) does not depend upon the parameters ${\bf k}$ and
$\mbox{$\boldmath \omega$}$, since they are now determined functionals of
$\mbox{$\boldmath \varphi$}(\theta,X)$
(in the vicinity of ${\cal M}^{\prime}$),
the values of which on the ${\cal M}^{\prime}$ coincide with the
corresponding parameters of the family. It can be easily seen that only
the functions from ${\cal M}^{\prime}$ satisfy to (\ref{fsv}).

The functionals ${\bf U}(X)$ and $\theta_{0}(X)$ in
$\Delta_{\delta}$ can be defined for instance by the following way:
let introduce in $\Delta_{\delta}$ $N$ functionals of the form
\begin{equation}
\label{an}
a^{\nu}[\mbox{$\boldmath \varphi$}](X) = {1 \over (2\pi)^{m}}
\int_{0}^{2\pi}\!\!\dots\int_{0}^{2\pi} A^{\nu}(\mbox{$\boldmath \varphi$}
(\theta,X),\mbox{$\boldmath \varphi$}_
{\theta^{\alpha}}(\theta,X),\dots)
d^{m}\theta
\end{equation}
with some functions $A^{\nu}$, such that on the ${\cal M}^{\prime}$
the values $a^{\nu}$ are functionally independent. On the
${\cal M}^{\prime}$ the values $a^{\nu}$, as can be easily seen, can be
expressed in terms of the values ${\bf U}(X)$ and do not depend upon
$\theta_{0}(X)$.
So that, dividing ${\cal M}^{\prime}$ into the "maps" in which
$|U_{(1)}^{\nu}(X) - U_{(2)}^{\nu}(X)| < \delta^{\prime}$,
we can express ${\bf U}(X)$ in terms of $a^{\nu}(X)$ in each of these maps
in the form $U^{\nu}(X) = f^{\nu}({\bf a}(X))$ and then, using the
definition (\ref{an}) of ${\bf a}(X)$ in $\Delta_{\delta}$,
we can extend $U^{\nu}(X)$ into the vicinity of each map
(this can be done at each $X$ independently).
Then, after the definition of the functionals $U^{\nu}(X)$ in
$\Delta_{\delta}$, let consider in $\Delta_{\delta}$ the functionals:
\begin{equation}
\label{vth}
\vartheta^{\alpha}(X) = {1 \over (2\pi)^{m}}\int_{0}^{2\pi}\dots\int_{0}^
{2\pi} \sum_{i} \varphi^{i}(\theta,X)
\Phi^{i}_{in\theta^{\alpha}}
(\theta,{\bf U}[\mbox{$\boldmath \varphi$}](X))
d^{m}\theta ,
\end{equation}
where $\mbox{$\boldmath \Phi$}_{in}(\theta,{\bf U})$
are introduced in (\ref{fms}) functions from ${\cal M}^{\prime}$.

If $\mbox{$\boldmath \varphi$}(\theta,X) \in
{\cal M}^{\prime}$ and $\theta_{0}(X) \equiv 0$, then
$\vartheta^{\alpha}(X) \equiv 0$ and, in the generic situation,
at small $\theta_{0}^{\alpha}(X)$ the values
$\{\theta_{0}^{\alpha}(X)\}$ can be expressed in terms of
$\{\vartheta^{\alpha}(X)\}$ on ${\cal M}^{\prime}$ in the form:
$\theta_{0}^{\alpha}(X) = \tau^{\alpha}(\mbox{$\boldmath \vartheta$}(X))$.
After that, by the same way, dividing, if it is necessary,
each of the maps described above into the parts in which:
$|\theta_{0(1)}^{\alpha}(X) - \theta_{0(2)}^{\alpha}(X)| < \delta^
{\prime\prime}$ (independently at each $X$),
and expressing by such formulas $\theta_{0}^{\alpha}(X)$
in terms of the $\mbox{$\boldmath \vartheta$}(X)$,
we can extend $\theta_{0}^{\alpha}(X)$
into $\Delta_{\delta}$ using the functionals
$\mbox{$\boldmath \vartheta$}(X)$.

It can be easily checked that after such definition of ${\bf U}(X)$ and
$\theta_{0}(X)$ in $\Delta_{\delta}$ we have for two functions
$\mbox{$\boldmath \varphi$}_{(1)}(\theta,X)$ and
$\mbox{$\boldmath \varphi$}_{(2)}(\theta,X)$ from
$\Delta_{\delta}$, satisfying the condition of type (\ref{del}), that is
$\|\mbox{$\boldmath \varphi$}_{(1)}(\theta,X) -
\mbox{$\boldmath \varphi$}_{(2)}(\theta,X)\|
< \delta$, where $\delta$
is small, the relation:
$|U^{\nu}_{(1)}(X) - U^{\nu}_{(2)}(X)| < C\delta , |\theta_{0(1)}^
{\alpha}(X) - \theta_{0(2)}^{\alpha}(X)| < C\delta$, where $C$ is
constant. The analogous relations will be 
also satisfied for the variational
derivatives of ${\bf U}(X)$ and $\theta_{0}(X)$.

We shall suppose that after the prolongation described above the
submanifold ${\cal M}^{\prime}$, given by the system (\ref{fsv}),
possesses the property of regularity, analogous to the regularity of
${\cal M}_{\mbox{$\boldmath \omega$},{\bf k}}$. Namely, it is clear
that the vectors:
$\mbox{$\boldmath \Phi$}_{\theta^{\alpha}}(\theta+
\theta_{0}(X),{\bf U}(X))$ and
$\mbox{$\boldmath \Phi$}_{U^{\nu}}(\theta+\theta_{0}(X),{\bf U}(X))$
(let us denote this set as
$\{{\tilde {\mbox{$\boldmath \xi$}}}^{(q)}_{[{\bf U},\theta_{0}]}(X),
q = 1,\dots,N+m \}$) lie at all $X$ at the kernel of linearized on the
function from ${\cal M}^{\prime}$ (characterized by the corresponding
values ${\bf U}(X)$ and
$\theta_{0}(X)$) functional
$F^{i}(\theta,X)[\mbox{$\boldmath \varphi$}]$,
introduced in (\ref{fsv}):
$$\delta F^{i}(\theta, X) =
{1 \over (2\pi)^{m}}\int_{0}^{2\pi}\!\!\dots\int_{0}^{2\pi}
{\tilde L}^{i}_{j[{\bf U}]}
(\theta + \theta_{0}(X),\theta^{\prime} + \theta_{0}(X),X)
\delta \varphi^{j}(\theta^{\prime},X) d^{m}\theta^{\prime} =$$
$$= {1 \over (2\pi)^{m}}\int_{0}^
{2\pi}\dots\int_{0}^{2\pi} [( \omega^
{\alpha}({\bf U}(X)) \delta^{i}_{j} \delta_{\theta^{\alpha}}(\theta -
\theta^{\prime})
- {\partial Q^{i} \over \partial \varphi^{j}}|_{{\bf k}=const}
\delta(\theta -\theta^{\prime}) -$$
$$- {\partial Q^{i} \over \partial \varphi^{j}_{\theta^{\alpha}}}
|_{{\bf k}=const} \delta_{\theta^{\alpha}}(\theta -\theta^{\prime}) -
{\partial Q^{i} \over \partial \varphi^{j}_{\theta^{\alpha}\theta^
{\beta}}}|_{{\bf k}=const} \delta_{\theta^{\alpha}\theta^
{\beta}}(\theta -\theta^{\prime}) - \dots) +$$
\begin{equation}
\label{fff}
+ ({\partial \omega^{\alpha} \over
\partial U^{\nu}}(X){\delta U^{\nu}(X) \over \delta
\varphi^{j}(\theta^
{\prime},X)}\varphi^{i}_{\theta^{\alpha}}(\theta,X) -
{\partial Q^{i} \over \partial U^{\nu}}(\theta,X)
{\delta U^{\nu}(X) \over \delta \varphi^{j}(\theta^
{\prime},X)})] \delta \varphi^{j}(\theta^
{\prime},X) d^{m}\theta^{\prime} .
\end{equation}

We shall assume that there is no other vectors possessing this property
and, besides that, at all points of ${\cal M}^{\prime}$
there exist at all $X$ exactly $N+m$ of "left eigen vectors"
${\tilde {\mbox{$\boldmath \kappa$}}}^{(q)}_
{[{\bf U},\theta_{0}]}(X)$
for the operator ${\hat {\tilde {\bf L}}}_{[{\bf U},\theta_{0}]}$
with zero eigen values, that is
such linearly independent at each $X$ $2\pi -$
periodic functions of $\theta$:
${\tilde \kappa}^{(q)}_{i [{\bf U}]}
(\theta + \theta_{0},X),\,\, i=1,\dots,n)$, that:
\begin{equation}
\label{tlsv}
{1 \over (2\pi)^{m}}\int_{0}^{2\pi}\!\!\dots\int_{0}^{2\pi}
{\tilde \kappa}^{(q)}_{i [{\bf U}]}
(\theta,X) {\tilde L}^{i}_{j[{\bf U}]}
(\theta,\theta^{\prime},X)
d^{m}\theta \equiv 0 .
\end{equation}

 Note that unlike
$\{{\tilde {\mbox{$\boldmath \xi$}}}^{(q)}_
{[{\bf U},\theta_{0}]}(X)\}$
the form of functions $\{{\tilde {\mbox{$\boldmath \kappa$}}}^{(q)}_
{[{\bf U},\theta_{0}]}(X)\}$
depends upon the manner of prolongation of ${\bf U}(X)$ and
$\theta_{0}(X)$ into the
$\Delta_{\delta}$, since ${\tilde {\mbox{$\boldmath \xi$}}}^{(q)}$
from the kernel of ${\hat {\tilde {\bf L}}}$ correspond to the variations
of functions $\mbox{$\boldmath \varphi$}(\theta,X)$ along the
${\cal M}^{\prime}$, on which ${\bf U}(X)$ and $\theta_{0}(X)$
are defined initially, whereas the
${\tilde {\mbox{$\boldmath \kappa$}}}^{(q)}(\theta,X)$
are connected with the image of the operator ${\hat {\tilde {\bf L}}}$,
for finding of which it is necessary to know the change of
${\bf U}(X)$ and $\theta_{0}(X)$
at all variations of $\mbox{$\boldmath \varphi$}(\theta,X)$.

With the aid of the constructions described above we can
introduce another system of coordinates in $\Delta_{\delta}$ instead
of the standard system, given by the
functionals $\varphi^{i}(\theta,X)$. Namely, we take as the
coordinates in $\Delta_{\delta}$ the functionals
$\{U^{\nu}(X), \theta_{0}^{\alpha}(X), \nu = 1,\dots,N,
\alpha = 1,\dots,m\}$, and
$$G^{i}_{[{\bf U},\theta_{0}]}[\mbox{$\boldmath \varphi$}](\theta,X) =
{1 \over (2\pi)^{m}}\int_{0}^{2\pi}\!\!\dots\int_{0}^{2\pi}
{\tilde L}^{i}_{j [{\bf U}]}(\theta + \theta_{0}(X),
\theta^{\prime} + \theta_{0}(X),X)
\times$$
\begin{equation}
\label{ccc}
\times(\varphi^{j}(\theta^{\prime},X) -
\Phi^{j}_{in}(\theta^{\prime} +
\theta_{0}(X),{\bf U}(X)) d^{m}\theta^{\prime} ,
\end{equation}
where ${\hat {\tilde L}}^{i}_{j [{\bf U}]}$ are functionals
from (\ref{fff}), appeared in the linearization of the constraints
$F^{i}(\theta,X)$ on the submanifold ${\cal M}^{\prime}$.

So that, the functions $\mbox{$\boldmath \varphi$}(\theta,X)$ from
$\Delta_{\delta}$ will be characterized by the set of $N+m$ smooth
functions ${\bf U}^{\nu}(X),
\theta_{0}^{\alpha}(X)$, and by the functions
$G^{i}(\theta,X), i = 1,\dots,n$,
taking the values at the space of $2\pi -$ periodic with respect to
$\theta$ functions, satisfying at each $X$ (and given ${\bf U}(X)$
and $\theta_{0}(X)$) to $N+m$ linear integral conditions of type:
\begin{equation}
\label{rassl}
{1 \over (2\pi)^{m}}\int_{0}^{2\pi}\!\!\dots\int_{0}^{2\pi}
{\tilde \kappa}^{(q)}_{i [{\bf U}]}
(\theta + \theta_{0}(X),X) G^{i}(\theta,X)
d^{m}\theta = 0
\end{equation}
(The conditions of such type are not customary in usual systems of
coordinates, however, they can be encountered in the theory of
stratifications).

Lemma 2.\newline
At small enough $\delta$ the values of functionals
$U^{\nu}(X), \theta_{0}^{\alpha}(X)$ and
$G^{i}(\theta,X)$ in $\Delta_{\delta}$, satisfying (\ref{rassl}),
give uniquely the values $\varphi^{i}(\theta,X)$.

Proof.\newline
Indeed, the system:
$${1 \over (2\pi)^{m}}\int_{0}^{2\pi}\!\!\dots\int_{0}^{2\pi}
{\tilde L}^{i}_{j[{\bf U}]}(\theta + \theta_{0}(X),
\theta^{\prime} + \theta_{0}(X),X)
\times$$
$$\times(\varphi^{j}(\theta^{\prime},X) -
\Phi^{j}_{in}(\theta^{\prime} +
\theta_{0}(X),{\bf U}(X))
d^{m}\theta^{\prime m} = G^{i}(\theta,X)$$
has, view (\ref{rassl}), a solution, defined
modulo the linear combination of the vectors
$\{\mbox{$\boldmath \xi$}^{(q)}_{[{\bf U},\theta_{0}]}(X)\}$,
tangential to ${\cal M}^{\prime}$ and corresponding to variations of
parameters ${\bf U}(X)$ and $\theta_{0}^{\alpha}(X)$; so that,
their coefficients in small enough $\Delta_{\delta}$ may be defined
by the values $U^{\nu}(X)$ and $\theta_{0}^{\alpha}(X)$.

We shall also need another system of coordinates in $\Delta_{\delta}$,
which connected with the system described above by the transformation
depending upon $\epsilon$.

Namely, consider the functionals:
\begin{equation}
\label{ab}
J^{\nu}(X) = {1 \over (2\pi)^{m}}\int_{0}^{2\pi}\!\!\dots\int_{0}^{2\pi}
{\cal P}^{\nu}(\mbox{$\boldmath \varphi$}(\theta,X),
\epsilon\mbox{$\boldmath \varphi$}_{X}(\theta,X),\dots)
d^{m}\theta ,\,\,\, \nu = 1,\dots,N ,
\end{equation}
and
\begin{equation}
\label{bc}
\theta_{0}^{*}(X) = \theta_{0}(X) -
{1 \over \epsilon}\int_{X_{0}}^{X}{\bf k}({\bf J}(X^{\prime}))dX^{\prime}
\end{equation}
(for some initial point $X_{0}$). As a system of coordinates
in the vicinity of ${\cal M}^{\prime}$ we take the functionals
$\{J^{\nu}(X), \theta_{0}^{*}(X)\}$, and also the set of constraints
$G^{i}(\theta,X)$ satisfying to (\ref{rassl}).

 The transition from $(U^{\nu}(X),\theta_{0}^{\alpha}(X))$ to
$(J^{\nu}(X),\theta_{0}^{*\alpha}(X))$,
as can be easily seen from (\ref{bc}),
is defined at $\epsilon \neq 0$. On the submanifold
${\cal M}^{\prime}$ (that is at $G^{q}(\theta,X) \equiv 0$)
at $\epsilon \rightarrow 0$ we have:
$$J^{\nu}(X) = {1 \over (2\pi)^{m}}\int_{0}^
{2\pi}\!\!\dots\int_{0}^{2\pi}
{\cal P}^{\nu}(\mbox{$\boldmath \Phi$}_{in}(\theta +
\theta_{0}(X),{\bf U}(X)),$$
$$\epsilon (\theta_{0 X}^{\alpha}\mbox{$\boldmath \Phi$}_
{in \theta^{\alpha}}
+ U^{\nu}_{X}\mbox{$\boldmath \Phi$}_
{in U^{\nu}}),\dots) d^{m}\theta =$$
$$= \sum_{k \geq 0}\epsilon^{k}J^{\nu}_{(k)}({\bf U},{\bf U}_
{X},\dots,{\bf U}_{kX},\theta_{0 X},
\dots,\theta_{0 kX}) ,$$
where $J^{\nu}_{(k)}$ represents the averaged with respect to
$\theta$ the $k$th Teilour term of the expansion of
${\cal P}^{\nu}$ at $\epsilon \rightarrow 0$
on the functions from ${\cal M}^{\prime}$. There is a finite number
of terms in the sum, there introduced the notations:
${\bf U}_{kX} \equiv \partial^{k}{\bf U} / \partial X^{k},
\theta_{0 kX} \equiv
\partial^{k}\theta_{0} / \partial X^{k}$.
The expression for $\theta_{0}^{*}(X)$ in terms of
${\bf U}(X)$ and $\theta_{0}(X)$ on the
${\cal M}^{\prime}$ can be obtained by the substitution of these
expansions for ${\bf J}(X)$ into (\ref{bc}).

We shall need also the inverse transformation from ${\bf J}(X)$ and
$\theta_{0}^{*}(X)$ to ${\bf U}(X),
\theta_{0}(X)$ at $G^{q}(\theta,X) \equiv 0$
(that is on the ${\cal M}^{\prime}$).
Let us note that the values
$J^{\nu}(X), \theta_{0}^{*\alpha}$
and $U^{\mu}(X)$ are connected on ${\cal M}^{\prime}$ by the
relations (the definition of $J^{\nu}(X)$):
$$J^{\nu}(X) = {1 \over (2\pi)^{m}}\int_{0}^{2\pi}\!\!\dots\int_{0}^{2\pi}
{\cal P}^{\nu}(\mbox{$\boldmath \Phi$}_{in}(\theta +
\theta_{0}^{*}(X) + {1 \over \epsilon}\int_
{X_{0}}^{X}{\bf k}({\bf J}(X^{\prime}))dX^{\prime} , {\bf U}(X)),$$
$$\epsilon \partial_{X} \mbox{$\boldmath \Phi$}_
{in}(\theta+ \theta_{0}^{*}(X) +
{1 \over \epsilon}\int_{X_{0}}^{X}{\bf k}
({\bf J}(X^{\prime}))dX^{\prime} ,
{\bf U}(X)),\dots) d^{m}\theta =$$
$$= {1 \over (2\pi)^{m}}\int_{0}^{2\pi}\!\!\dots\int_{0}^{2\pi}
{\cal P}^{\nu}(\mbox{$\boldmath \Phi$}_{in}(\theta +
\theta_{0}^{*}(X) +
{1 \over \epsilon}\int_{X_{0}}^{X}{\bf k}
({\bf J}(X^{\prime}))dX^{\prime} ,
{\bf U}(X)),$$
$$k^{\alpha}({\bf J})
\partial_{\theta^{\alpha}} \mbox{$\boldmath \Phi$}_{in}
(\theta + \theta_{0}^{*}(X) +
{1 \over \epsilon}\int_{X_{0}}^{X}{\bf k}
({\bf J}(X^{\prime}))dX^{\prime} ,
{\bf U}(X)),\dots) d^{m}\theta +$$
$$+ \sum_{k \geq 1} \epsilon^{k} {1 \over (2\pi)^{m}}\int_{0}^
{2\pi}\dots\int_{0}^{2\pi} {\cal P}^{\nu}_
{(k)}(\mbox{$\boldmath \Phi$}_{in}
(\dots),\dots) d^{m}\theta , $$
- where ${\cal P}^{\nu}_{(k)}(\mbox{$\boldmath \Phi$}_
{in}(\dots),\dots)$
are local functionals depending upon
$\mbox{$\boldmath \Phi$}_{in}(\theta +\theta_{0}^{*}(X) +
{1 \over \epsilon}\int_{X_{0}}^{X}{\bf k}
({\bf J}(X^{\prime}))dX^{\prime} ,
{\bf U}(X))$ and their derivatives with respect to
$U^{\nu}$ and $\theta^{\alpha}$ with the coefficients of type:
${\bf U}_{X}(X), {\bf U}_{XX}(X),\dots,
{\bf k}({\bf J}),\partial_{X}{\bf k}({\bf J}),\linebreak
\partial_{X}^{2}{\bf k}({\bf J}),\dots$, and
$\theta_{0X}^{*}(X),\theta_{0XX}^{*}(X),\dots$,
given by the collecting together these terms, having the general
multiplier $\epsilon^{k}$. The term, corresponding to the zero power
of $\epsilon$, is written separately.

After the integration with respect to $\theta$, which removes the
singular at $\epsilon \rightarrow 0$ phase shift $\theta_{0}$
in the argument $\mbox{$\boldmath \Phi$}_{in}$,
we obtain on ${\cal M}^{\prime}$:
\begin{equation}
\label{cd}
J^{\nu}(X) = \zeta^{\nu}({\bf J},{\bf U}) + \sum_{k \geq 1} \epsilon^{k}
\zeta^{\nu}_{(k)}({\bf U},{\bf U}_{X},\dots,{\bf U}_{kX},{\bf J},
{\bf J}_{X},\dots,{\bf J}_{kX},
\theta_{0X}^{*},\dots,
\theta_{0kX}^{*}) .
\end{equation}
The sum in (\ref{cd}) contains the finite number of terms, functions
$\zeta^{\nu}_{(k)}$ and $\zeta^{\nu}$ are integrated with respect to
$\theta$ functions ${\cal P}^{\nu}_{(k)}$ and
${\cal P}^{\nu}$ respectively. Since
$$\zeta^{\nu}({\bf J},{\bf U}) = {1 \over (2\pi)^{m}}\int_{0}^
{2\pi}\dots\int_{0}^{2\pi} {\cal P}^{\nu}
(\mbox{$\boldmath \Phi$}_{in}(\theta,{\bf U}(X)),
k^{\alpha}({\bf J}) \partial_{\theta^{\alpha}}
\mbox{$\boldmath \Phi$}_{in}(\theta,{\bf U}(X)),$$
$$k^{\alpha}({\bf J})k^{\beta}({\bf J}) \partial_{\theta^{\alpha}}
\partial_{\theta^{\beta}} \mbox{$\boldmath \Phi$}_{in}
(\theta,{\bf U}(X)),\dots)
d^{m}\theta $$
(in the integration with respect to $\theta$ we omitted
the unessential phase shift $\mbox{$\boldmath \Phi$}_{in}$)
we obtain that the system:
\begin{equation}
\label{de}
J^{\nu}(X) = \zeta^{\nu}({\bf J}(X),{\bf U}(X))
\end{equation}
is satisfied by the solution $J^{\nu}(X) \equiv U^{\nu}(X)$,
according to the definition (\ref{parameters}) of parameters ${\bf U}(X)$
on the family ${\cal M}^{\prime}$. Since we suppose that the system
(\ref{de})
is of general form, we shall assume that it is equivalent to the system
$J^{\nu}(X) = U^{\nu}(X)$.

Taking this into account, we can resolve the system (\ref{cd}) by the
iterations, taking on the initial step
$U^{\nu}(X) = J^{\nu}(X)$; for $U^{\nu}(X)$ we shall obtain
the expression of the form:
\begin{equation}
\label{ef}
U^{\nu}(X) = J^{\nu}(X) + \sum_{k \geq 1} \epsilon^{k} U^{\nu}_{(k)}
({\bf J},{\bf J}_{X},\dots,{\bf J}_{kX},\theta_
{0X}^{*},\dots,\theta_{0kX}^{*}) .
\end{equation}
The substitution of (\ref{ef}) into (\ref{cd}), under the condition
of nonsingularity of matrix
$\|{\partial \zeta^{\nu}({\bf J},{\bf U}) \over \partial U^{\mu}}\||_
{{\bf U} = {\bf J}}$, will uniquely define the functions $U^{\nu}_{(k)}$.
The value $\theta_{0}(X)$ is expressed in terms of
${\bf J}(X)$ and $\theta_{0}^{*}(X)$
according to the formula (\ref{bc}).

Later it will be convenient to write the expressions of type
$\theta_{0}^{\alpha}[{\bf J},\theta_{0}^{*},\epsilon]$
and $U^{\nu}[{\bf J},\theta_{0}^{*},\epsilon]$,
assuming the expressions (\ref{bc}) and (\ref{ef})
(not necessary on the ${\cal M}^{\prime}$).

The functionals (\ref{ab}) are well defined on the full space
of functions $\mbox{$\boldmath \varphi$}(\theta,X)$
and the Hamiltonian flows, generated with the aid of (\ref{esk})
by the integrals:
\begin{equation}
\label{qin}
\int q(X) J^{\nu}(X) dX = \int q(X)
{1 \over (2\pi)^{m}}\int_{0}^{2\pi}\!\!\dots\int_{0}^{2\pi}
{\cal P}^{\nu}(\mbox{$\boldmath \varphi$},
\epsilon\mbox{$\boldmath \varphi$}_{X},\dots) d^{m}\theta ,
\end{equation}
have for all smooth functions $q(X)$ the form:
\begin{equation}
\label{qfl}
\varphi^{i}_{t} = q(X)Q^{i}_{\nu}(\mbox{$\boldmath \varphi$},
\epsilon\mbox{$\boldmath\varphi$}_{X},\dots) + \epsilon q_{X}
{\tilde Q}^{i}_{\nu}(\mbox{$\boldmath \varphi$},
\epsilon\mbox{$\boldmath\varphi$}_{X},\dots) +
\epsilon^{2} q_{XX}{\tilde {\tilde Q}}^{i}_
{\nu}(\mbox{$\boldmath \varphi$},
\epsilon\mbox{$\boldmath\varphi$}_{X},\dots) + \dots ,
\end{equation}
where $Q^{i}_{\nu}$ are introduced in (\ref{fluxes}) flows generated
by the functionals $I^{\nu}$,
${\tilde Q}^{i}_{\nu}, \linebreak {\tilde {\tilde Q}}^{i}_{\nu},\dots$
are some functions. (The derivatives of $q(X)$ with respect to $X$
arise in the calculation of the variational derivative and also
as a result of action of bracket (\ref{esk}). As can be easily seen,
each differentiation of $q(X)$ with respect to $X$ appears with the
multiplier $\epsilon$.)

\begin{center}
{\bf 5. The restriction of the Poisson bracket on the submanifold
corresponding to the full family of m-phase solutions.}
\end{center}

For the Dirac restriction of bracket (\ref{esk}) on the
submanifold ${\cal M}^{\prime}$
in the coordinates ${\bf J}(X),\theta_{0}^{*}(X)$ and
${\bf G}(\theta,X) = (G^{i}(\theta,X))$
we shall need their Poisson brackets with each other
(on the ${\cal M}^{\prime}$).

We shall begin with the brackets of type: $\{J^{\nu}(X),J^{\mu}(Y)\}$.
View (\ref{skob}), these brackets have the form:
$$\{J^{\nu}(X),J^{\mu}(Y)\} =$$
$$= {1 \over (2\pi)^{2m}}\int_{0}^{2\pi}\dots\int_
{0}^{2\pi} \sum_{k \geq 0} A_{k}^{\nu\mu}
(\mbox{$\boldmath \varphi$},\epsilon\mbox{$\boldmath\varphi$}_{X},\dots)
\epsilon^{k} \delta^{(k)}(X-Y) \delta (\theta -
\theta^{\prime}) d^{m}\theta d^{m}\theta^{\prime} =$$
\begin{equation}
\label{fg}
= {1 \over (2\pi)^{m}}\int_{0}^{2\pi}\!\!\dots\int_{0}^{2\pi}
\sum_{k \geq 0}A_{k}^{\nu\mu}(\mbox{$\boldmath \varphi$},
\epsilon\mbox{$\boldmath\varphi$}_{X},\dots)
\epsilon^{k} \delta^{(k)}(X-Y)
d^{m}\theta ,
\end{equation}
where view (\ref{aq}):
\begin{equation}
\label{gh}
A_{0}^{\nu\mu}(\mbox{$\boldmath \varphi$},
\epsilon\mbox{$\boldmath \varphi$}_{X},
\epsilon^{2}\mbox{$\boldmath \varphi$}_{XX},\dots)
\equiv \epsilon \partial_{X}Q^{\nu\mu}(\mbox{$\boldmath \varphi$},
\epsilon\mbox{$\boldmath \varphi$}_{x},
\epsilon^{2}\mbox{$\boldmath \varphi$}_{XX},\dots) .
\end{equation}
At the points of the submanifold ${\cal M}^{\prime}$ the function
$\mbox{$\boldmath \varphi$}(\theta,X)$ has the form:
$\varphi^{i}(\theta,X) = \Phi^{i}_{in}(\theta + \theta_{0}^{*}(X) +
{1 \over \epsilon}\int_{X_{0}}^{X}{\bf k}({\bf J}(X^{\prime}))dX^{\prime},
{\bf U}[{\bf J},\theta_{0}^{*},\epsilon](X))$
and after the substitution of it into (\ref{fg}) we can obtain
the brackets
$\{J^{\nu}(X),J^{\mu}(Y)\}$ at the points of ${\cal M}^{\prime}$
(characterized by coordinates
${\bf J}(X),\theta_{0}^{*}(X))$ in form of regular at
$\epsilon \rightarrow 0$ series, since the integration with respect to
$\theta$ removes the irregularity
at $\epsilon \rightarrow 0$ in the argument of function
$\mbox{$\boldmath \Phi$}_{in}$. View (\ref{gh}), it can be easily
seen that the zero terms are absent in these series, so that
they can be written in the form:
$$\{J^{\nu}(X),J^{\mu}(Y)\}|_{{\bf G}(\theta,X) = 0} =$$
\begin{equation}
\label{jj}
= \epsilon
\left(\langle A_{1}^{\nu\mu}\rangle({\bf J}(X))
\delta^{\prime}(X-Y) +
{\partial\langle Q^{\nu\mu}\rangle({\bf J}(X))
\over \partial X} \delta(X-Y)
\right) + \epsilon^{2}{\cal J}^{\nu\mu}[{\bf J},\theta_
{0}^{*},\epsilon] ,
\end{equation}
where ${\cal J}^{\nu\mu}$ are regular at $\epsilon \rightarrow 0$
functionals of ${\bf J}$ and $\theta_{0}^{*}$.
The functionals of this type will arise later and have the form
of regular at $\epsilon \rightarrow 0$ asymptotic series of type:
$$C[{\bf J},\theta^{*}_{0},\epsilon] = \sum_{k\geq0}
C_{(k)}({\bf J},{\bf J}_{X},\dots,{\bf J}_{kX},\theta^{*}_{0X},
\dots,\theta^{*}_{0kX}) \epsilon^{k}.$$

Here we used the relation
(\ref{ef}) giving ${\bf U}(X)$ in terms of ${\bf J}(X)$ and
$\theta_{0}^{*}(X)$ at the points of
${\cal M}^{\prime}$, according to which the arguments $U^{\nu}(X)$,
on which depend the values $\langle A_{1}^{\nu\mu}\rangle$
and $\langle Q^{\nu\mu}\rangle$ on ${\cal M}^{\prime}$, are replaced
modulo the terms of higher order of $\epsilon$ by $J^{\nu}(X)$,
$\langle\dots\rangle$ means, as previously,
the averaging on the family of m-phase solutions of (\ref{system}),
defined by the formula (\ref{us}).

The evolution of the densities of conservation laws
${\cal P}^{\nu}(\mbox{$\boldmath \varphi$},
\epsilon\mbox{$\boldmath\varphi$}_{X},\dots)$ according to the system
(\ref{qfl}), generated by the functional $\int q(X) J^{\mu}(X) dX$,
as can be easily seen from (\ref{skob}), has the form:
\begin{equation}
\label{hij}
{\cal P}^{\nu}_{\tau^{\mu}}(X) = \sum_{k \geq 0}
A_{k}^{\nu\mu}(\mbox{$\boldmath \varphi$},
\epsilon\mbox{$\boldmath \varphi$}_{XX},
\epsilon^{2}\mbox{$\boldmath \varphi$}_{XX},\dots)
\epsilon^{k} q_{kX}(X) .
\end{equation}
The evolution of $J^{\nu}(X)$ at the points of ${\cal M}^{\prime}$,
characterized by the coordinates ${\bf J}(X),\theta_{0}^{*}(X)$,
will be written in the form:
$$J^{\nu}_{\tau^{\mu}}(X) = \epsilon
\left(\langle A_{1}^{\nu\mu}\rangle({\bf J}(X))
q_{X}(X) + {\partial\langle
Q^{\nu\mu}\rangle({\bf J}(X)) \over \partial X}
q(X)\right) +$$
\begin{equation}
\label{jk}
+ \epsilon^{2} f^{\nu\mu}[{\bf J},\theta_
{0}^{*},q(X),\epsilon] ,\,\,\, \nu,\mu = 1,\dots,N ,
\end{equation}
where $f^{\nu\mu}$ are regular at $\epsilon \rightarrow 0$ asymptotic
series with respect to $\epsilon^{k}$, the coefficients of which are
local functionals of $q(X), {\bf J}(X)$ and $\theta_{0}^{*}(X)$.

Lemma 3. \newline
From the system (\ref{jk}) ($\nu = 1,\dots,N, \mu$ - is fixed),
giving the evolution of ${\bf J}(X)$ according to the system (\ref{qfl}),
at the points of ${\cal M}^{\prime}$ follow the
relations:
\begin{equation}
\label{oxk}
k^{\alpha}_{\tau^{\mu}} ({\bf J}(X)) =
\epsilon \left(q(X) \omega^{\alpha}_
{\mu}({\bf J}(X))\right)_{X} +
\epsilon^{2} {\tilde k}^{\alpha}_{\mu}
[{\bf J},\theta_{0}^{*},q(X),\epsilon] ,
\end{equation}
- where $\mbox{$\boldmath \omega$}_{\nu} =
(\omega^{1}_{\nu},\dots,\omega^{m}_{\nu})$
are introduced in (\ref{fom}) functions,
${\tilde k}^{\alpha}_{\mu}$
are regular at $\epsilon \rightarrow 0$ functionals of
${\bf J},\theta_{0}^{*}$
and $q(X)$ (linear with respect to $q(X)$).

Proof. \newline
System (\ref{qfl}) admits the construction of the family of solutions
similar to the described in Whitham's method ones, that is the
family of solutions of type:
\begin{equation}
\label{kl}
\mbox{$\boldmath \varphi$} = \mbox{$\boldmath \varphi$}
\left({\mbox{$\boldmath \sigma$}(X,T) \over \epsilon} + \theta,
X,T,\epsilon\right) = \sum_{k \geq 0}\mbox{$\boldmath \varphi$}_{k}
\left({\mbox{$\boldmath \sigma$}(X,T)
\over \epsilon} + \theta, X,T\right) \epsilon^{k} ,
\end{equation}
$T = \epsilon \tau^{\mu}$.

By the substitution of (\ref{kl}) into (\ref{qfl}) we obtain in the zero
order that $\mbox{$\boldmath \varphi$}_{0}(\theta,X,T)$
at each $T$ is the function from the ${\cal M}^{\prime}$, since
it satisfies to one of the systems (\ref{fom}) (we assume here
that $I^{\mu}$ is not the momentum operator or annihilator of bracket
(\ref{bracket}), for which the statement of Lemma is evident), besides
that:
\begin{equation}
\label{lm}
\sigma^{\alpha}_{T} = q(X)\omega^{\alpha}_{\mu}({\bf U}(X)),\,\,\,\,\,
\sigma^{\alpha}_{X} = k^{\alpha}({\bf U}(X)) .
\end{equation}
We assume, as it was declared previously, that the systems
(\ref{fluxes}), generated by the funcionals $I^{\mu}$ (except the
annihilators and momentum operator), and corresponding to them
submanifolds ${\cal M}_{\mbox{$\boldmath \omega$}_{\mu},{\bf k}}$
satisfy to the nondegeneracy conditions $(A)$ and $(B)$,
which permit to construct the asymptotic series (\ref{kl}).
The conditions of the solvability of system on
$\mbox{$\boldmath \varphi$}_{1}(\theta,X,T)$, that is $N-m$
equations of orthogonality of the discrepancy $f_{1}$ to the left
eigen vectors of the corresponding operator ${\hat {\bf L}}_
{[{\bf U}(X),\theta_{0}(X)]}$, together with the equations
\begin{equation}
\label{mn}
{\bf k}({\bf U})_{T} =
(q(X)\mbox{$\boldmath \omega$}_{\mu}({\bf U}))_{X} ,
\end{equation}
following from (\ref{lm}), give, as previously,
the system of equations on ${\bf U}(X)$ and $\theta_{0}(X)$.

It can be easily seen (view (\ref{lm})), that the system on
$\mbox{$\boldmath \Phi$}_{(1)}(\theta,X,T)$
has now the form:
$$q(X) \left({\hat {\bf L}}^{\mu}_{[{\bf U}(X),\theta_
{0}(X)]}\mbox{$\boldmath \Phi$}_{1}\right) (\theta,X,T)
= q(X) [\sum_{\bf n} \alpha^{i(\mu)}_
{j({\bf n})}(\mbox{$\boldmath \sigma$}_{X},
\mbox{$\boldmath \Phi$}_{(0)},
\mbox{$\boldmath \Phi$}_{(0)\theta^{\alpha}},\dots) \times$$
$$\times\left( \mbox{$\boldmath \Phi$}_{(0){\bf n}\theta,
U^{\nu}} U^{\nu}_{X} + \mbox{$\boldmath \Phi$}_
{(0){\bf n}\theta,\theta^{\alpha}} \theta^
{\alpha}_{0X} \right) + \beta^{i(\mu)}_{\alpha}(\mbox{$\boldmath
\sigma$}_{X},\mbox{$\boldmath \Phi$}_{(0)},\mbox{$\boldmath \Phi$}_
{(0)\theta^{\alpha}},\dots) \mbox{$\boldmath \sigma$}_{XX}] +$$
$$+ q_{X}(X){\tilde Q}^{i}_{\mu}
(\mbox{$\boldmath \Phi$}_{(0)},\mbox{$\boldmath \sigma$}_
{X}\mbox{$\boldmath \Phi$}_{(0)\theta^{\alpha}},
\dots) - \left(\mbox{$\boldmath \Phi$}^{i}_{(0)U^{\nu}} U^{\nu}_{T} +
\mbox{$\boldmath \Phi$}^
{i}_{(0)\theta^{\alpha}} \theta^{\alpha}_{0T}\right) ,$$
all the notations are the same that the notations in (\ref{dd}),
the values $\mbox{$\boldmath \sigma$}_{T}$ and
$\mbox{$\boldmath \sigma$}_{X}$
are connected with the parameters ${\bf U}(X)$, corresponding to
$\mbox{$\boldmath \Phi$}_{(0)}(\theta,X,T)$,
by the relations (\ref{lm}), the functions
$\alpha^{i(\mu)}_{j({\bf n})}$ and $\beta^{j(\mu)}_{\alpha}$ are
the same that in Whitham's method for the evolution system,
generated by the operator $I^{\nu}$. The last property permits to
conclude that, similar to Whitham's situation, the conditions of
orthogonality of discrepancy $f_{1}$ to the left eigen vectors
of ${\hat {\bf L}}$ do not impose any restriction on the parameters
$\theta_{0}(X,T)$ and have a form:
\begin{equation}
\label{no}
q(X) A^{\zeta}_{\nu}({\bf U}) U^{\nu}_{X} + B^{\zeta}_{\nu}({\bf U})
U^{\nu}_{T} + q_{X}(X) C^{\zeta}({\bf U}) = 0 ,\,\,\, \zeta = 1,
\dots,N-m .
\end{equation}
Together with the equations (\ref{mn}), the system  (\ref{no}) gives $N$
equations on the functions ${\bf U}(X)$.

At given initial dates the system (\ref{mn})-(\ref{no})
has the smooth solution at $T < T_{0}$ (up to the moment of
destruction of waves), and in this region, as in Whitham's method,
we can, using the function
$\mbox{$\boldmath \varphi$}_{(0)}(\theta,X,T)$ from
${\cal M}^{\prime}$, characterized by the values
${\bf U}(X,T)$ and $\theta_{0}(X,T)$, construct uniquely
the next terms
$\mbox{$\boldmath \varphi$}_{(k)}(\theta,X,T)$ of the series
(\ref{kl}), which will be the functionals of
${\bf U}(X,T)$ and $\theta_{0}(X,T)$.
Taking this into account, and using the
definition of ${\bf J}(X)$ (\ref{ab})
and the parameters ${\bf U}(X)$ (\ref{parameters}),
we can obtain on the family (\ref{kl})
of solutions of (\ref{qfl}) the relations:
\begin{equation}
\label{op}
J^{\nu}(X,T) = U^{\nu}(X,T) +
\sum_{k\geq1}\epsilon^{k} {\tilde J}^{\nu}_
{(k)}[{\bf U},\theta_{0}] ,
\end{equation}
- where ${\bf U}(X,T)$ and $\theta_{0}(X,T)$ are parameters,
connected with the function
$\mbox{$\boldmath \varphi$}_{(0)}(\theta,X,T)$
from ${\cal M}^{\prime}$, ${\bf J}(X)$ are
values of the corresponding functionals on the solutions
(\ref{kl}), corresponding to the given ${\bf U}(X,T)$ and
$\theta_{0}(X,T)$.

After the substitution of
$\mbox{$\boldmath \varphi$}(\theta,X,T)$ in form of
(\ref{kl}) into the expression
(\ref{hij}) and the integration with respect to $\theta$
we obtain the expression for the evolution
of ${\bf J}(X)$ on the solutions
(\ref{kl}), which, as can be easily seen, has the form:
\begin{equation}
\label{pr}
J^{\nu}_{\tau^{\mu}}(X) = \epsilon \left( q_{X}(X)\langle
A^{\nu\mu}\rangle ({\bf U}(X))
 + q(X) \partial_{X}\langle Q^{\nu\mu}\rangle({\bf U}(X)) \right) +
\epsilon^{2} {\tilde f}^{\nu\mu}[{\bf U},\theta_{0},\epsilon] .
\end{equation}
Comparing (\ref{pr}) and (\ref{op}) we can conclude that the system
$$U^{\nu}_{T} = q_{X}(X)\langle A^{\nu\mu}_{1}\rangle({\bf U}(X)) +
q(X) \partial_{X}\langle Q^{\nu\mu}\rangle({\bf U}(X)) , \,\,\,
\,\,\,\,\,(T = \epsilon \tau^{\mu}), \,\,\, -$$
is equivalent to (\ref{mn},\ref{no}), and so that, the relations
(\ref{oxk}) follow from (\ref{jk}).
The Lemma is proved.

Corollary. \newline
On the functions of family ${\cal M}^{\prime}$, characterized by
coordinates ${\bf J}(X),\theta_{0}^{*}(X)$ at
$\epsilon \rightarrow 0$, take place the following relations:
$$\{k^{\alpha}(X),J^{\mu}(Y)\} =
\epsilon \left(\omega^{\alpha}_{\mu}({\bf J}(X))
\delta^{\prime}(X-Y) + {\partial \omega^{\alpha}_{\mu}({\bf J}(X))
\over \partial X} \delta (X-Y)\right) +$$
\begin{equation}
\label{kjk}
+ \epsilon^{2}k^{\alpha\mu}
[{\bf J},\theta_{0}^{*},\epsilon] , -
\end{equation}
where $k^{\alpha\mu}$ are regular at $\epsilon \rightarrow 0$
functionals of ${\bf J}(X)$ and $\theta_{0}^{*}(X)$.

Lemma 4.\newline
Under the formulated previously assumption that the number of parameters
$U^{\nu}$ is equal to $2m+p$, where $p$ is the number of the
annihilators of the bracket (\ref{bracket}), the following conditions
take place:
\begin{equation}
\label{rs}
\sum_{\nu = 1}^{N} {\partial k^{\alpha}({\bf U}) \over U^{\nu}}
\omega^{\beta}_{\nu}({\bf U}) \equiv 0 , \,\,\, \forall \alpha,\beta .
\end{equation}

Proof.\newline
As was formulated previously, any of $m+1$ flows, generated by
$m+1$ integrals from the set (\ref{laws}), are linearly dependent
on the family of m - phase solutions of system (\ref{system}),
and for some $\lambda^{\nu}_{s}({\bf U}), \,\,\,
\nu = 1,\dots,N, s = 1,\dots,N-m$ we have on ${\cal M}^{\prime}$ 
$(N-m)$ independent relations:
\begin{equation}
\label{st}
\sum_{\nu = 1}^{N} \lambda^{\nu}_{s}({\bf U})
\omega^{\alpha}_{\nu}({\bf U}) = 0 ,\,\,\,\,\, \alpha = 1,\dots,m ,
\end{equation}
- where $\omega^{\alpha}_{\nu}({\bf U})$ are introduced in
(\ref{fom}) frequencies, corresponding to the integrals $I^{\nu}$.
It follows from this that on m - phase solutions of
(\ref{system}), characterized by the parameters ${\bf U}(X)$,
for some functions $\mu^{q}_{s}({\bf U})$ take place the relations:
\begin{equation}
\label{tu}
\sum_{\nu = 1}^{N} \lambda^{\nu}_{s}({\bf U})
\delta I[\mbox{$\boldmath \varphi$}] =
\sum_{q = 1}^{p} \mu^{q}_{s}({\bf U},\theta_{0})
\delta N_{q} , \,\,\,\,\,-
\end{equation}
where $N_{q}$ are annihilators of bracket (\ref{bracket}). Let us denote
the values of the annihilators $N_{q}$ on m - phase solutions of
(\ref{system}) by $n_{q}({\bf U})$ (the dependence of $\theta_{0}$
is absent because of the commutation of $N_{q}$ with $I^{\nu}$, generating
the linear dependence of the initial phases upon the time).
The values of the functionals $I^{\nu}$ on m - phase solutions
of (\ref{system}) are the parameters $U^{\nu}$, and for infinitely
small shift on the ${\cal M}$ along the direction ${\vec \xi}$
(in the space of parameters on the ${\cal M}$), tangential to the
submanifold ${\bf k}({\bf U}) = const, ({\bf k} = (k^{1},\dots,k^{m}))$,
we obtain the uniformly bounded variations of functions
$\varphi^{i}(x)$, so that, view (\ref{tu}), we obtain on ${\cal M}$:
$$\sum_{\nu = 1}^{N} \lambda^{\nu}_{s}({\bf U}) dU^{\nu} -
\sum_{q=1}^{p} \mu^{q}_{s}({\bf U},\theta_{0}) dn_{q}({\bf U}) = 0,$$
if $dk^{\alpha} = 0$.

 From this we can conclude that $\mu({\bf U},\theta_{0}) =
\mu({\bf U})$ (the dependence upon
$\theta_{0}$ is absent), and for some functions
$\tau^{\alpha}_{s}({\bf U}), \alpha = 1,\dots,m, s = 1,\dots,N-m,$
take place the relations:
\begin{equation}
\label{uv}
\sum_{\nu = 1}^{N} \lambda_{s}^{\nu}({\bf U}) dU^{\nu} =
\sum_{q=1}^{p} \mu^{q}_{s}({\bf U}) dn_{q}({\bf U}) + \sum_{\alpha = 1}^{m}
\tau^{\alpha}_{s}({\bf U}) dk^{\alpha}({\bf U}) .
\end{equation}
After the expression of $N-m = m+p$ differentials $dn_{q}({\bf U})$
and $dk^{\alpha}({\bf U})$ from
$N-m$ equations (\ref{uv}) (we assume that in the generic situation
this can be done) we obtain the relations:
\begin{equation}
\label{vw}
dk^{\alpha}({\bf U}) = \sum_{s=1}^{N-m} a^{\alpha s}({\bf U})
\sum_{\nu = 1}^{N}
\lambda^{\nu}_{s}({\bf U}) dU^{\nu},
\end{equation}
\begin{equation}
\label{wx}
dn_{q}({\bf U}) = \sum_{s=1}^{N-m} b^{s}_{q}({\bf U}) \sum_{\nu = 1}^{N}
\lambda^{\nu}_{s}({\bf U}) dU^{\nu}
\end{equation}
for some functions $a^{\alpha s}({\bf U})$ and $b^{s}_{q}({\bf U})$.
After that the statement of the Lemma follows immediately
from (\ref{st}).
Lemma is proved.

It can be easily seen that, under the conditions formulated in Lemma 4,
for the functions $n_{q}({\bf U})$, view (\ref{st}) and (\ref{wx}),
also take place the relations similar to (\ref{rs}), that is:

Lemma 5. \newline
Under the formulated in Lemma 4 conditions: $N = 2m+p$, where $p$ is the
number of the annihilators of bracket (\ref{bracket}), on the submanifold
${\cal M}$, connected with the full family of m - phase
solutions of (\ref{system}), take place the relations:
\begin{equation}
\label{xy}
\sum_{\nu = 1}^{N} {\partial n_{q}({\bf U}) \over \partial U^{\nu}}
\omega^{\beta}_{\nu} \equiv 0 .
\end{equation}

Corollary. \newline
On the submanifold ${\cal M}^{\prime}$ with the coordinates
$({\bf J}(X),\theta_{0}^{*}(X))$
take place the relations:
\begin{equation}
\label{az}
\{k^{\alpha}({\bf J}(X)),k^{\beta}({\bf J}(Y))\} = \epsilon^{2}
\rho^{\alpha\beta}[{\bf J},\theta_{0}^{*},\epsilon] ,
\end{equation}
\begin{equation}
\label{bz}
\{k^{\alpha}({\bf J}(X)),n_{q}({\bf J}(Y))\} = \epsilon^{2}
\upsilon^{\alpha}_{q}[{\bf J},\theta_{0}^{*},\epsilon] , -
\end{equation}
where $\rho^{\alpha\beta}, \upsilon^{\alpha}_{q}$ are regular at
$\epsilon \rightarrow 0$ functionals of ${\bf J}(X)$ and
$\theta_{0}^{*}(X)$.

Indeed, using (\ref{kjk}) and (\ref{rs}), we obtain:
$$\{k^{\alpha}({\bf J}(X)),k^{\beta}({\bf J}(Y))\} =
\epsilon (\sum_{\nu = 1}^{N}
\omega^{\alpha}_{\nu}({\bf J}(X)){\partial k^{\beta}({\bf J}) \over
\partial J^{\nu}}(X) \delta^{\prime}(X-Y) +$$
$$+ \left(\omega^{\alpha}_
{\nu}({\bf J}(X)){\partial k^{\beta}({\bf J}) \over
\partial J^{\nu}}(X)\right)_{X} \delta(X-Y)) +
\epsilon^{2}k^{\alpha}_{\nu}[{\bf J},\theta_{0}^{*},\epsilon]
{\partial k^{\beta}({\bf J}) \over \partial J^{\nu}}(Y) =$$
$$= \epsilon^{2}k^{\alpha}_{\nu}[{\bf J},\theta_{0}^{*},\epsilon]
{\partial k^{\beta}({\bf J}) \over \partial J^{\nu}}(Y) .$$
The relation (\ref{bz}) is proved by the same way.

Now we shall calculate the brackets $\{J^{\nu}(X),\theta^
{*\alpha}_{0}(Y)\}$ and $\{\theta^{*\alpha}_{0}(X),\theta^{\beta*}_
{0}(Y)\}$ on ${\cal M}^{\prime}$. As was mentioned above,
the functionals of type $\int q(X)J^{\nu}(X)dX$
generate the local flows of form (\ref{qfl}). The corresponding
evolution of the densities
$A^{\nu}(\mbox{$\boldmath \varphi$},\mbox{$\boldmath \varphi$}_
{\theta^{\alpha}},\dots)$ of integrals $a^{\nu}(X)$,
introduced in (\ref{an}) for the definition of coordinates $U^{\nu}(X)$
in the vicinity of ${\cal M}^{\prime}$, also have the form similar to
(\ref{qfl}), that is
$$A^{\nu}_{\tau^{\mu}}(\theta,X) =
q(X)S^{\nu}_{\mu}(\mbox{$\boldmath \varphi$},
\mbox{$\boldmath \varphi$}_{\theta^{\alpha}},\dots,
\epsilon\mbox{$\boldmath \varphi$}_{X},
\epsilon\mbox{$\boldmath \varphi$}_{\theta^{\alpha}X},\dots) +$$
$$+ \epsilon q_{X}{\tilde S}^{\nu}_{\mu}(\mbox{$\boldmath \varphi$},
\theta_{\theta^{\alpha}},\dots,
\epsilon\mbox{$\boldmath \varphi$}_{X},
\epsilon\mbox{$\boldmath \varphi$}_{\theta^{\alpha}X},\dots)
+ \dots \,\,\, .$$

After the substitution of functions
$\mbox{$\boldmath \varphi$}(\theta,X)$
in the form:
\begin{equation}
\label{cz}
\mbox{$\boldmath \varphi$}(\theta,X) =
\mbox{$\boldmath \Phi$}_{in}(\theta +
\theta_{0}^{*}(X) + {1 \over \epsilon}\int_{X_{0}}^
{X}{\bf k}({\bf J}(X^{\prime}))dX^{\prime},
{\bf U}[{\bf J},\theta_{0}^{*},\epsilon])
\end{equation}
and integration with respect to $\theta$,
we obtain the expression for the evolution of $a^{\nu}(X)$ on the
functions from ${\cal M}^{\prime}$,
characterized by the coordinates ${\bf J}(X)$ and $\theta_{0}^{*}(X)$,
in the form:
$$a^{\nu}_{\tau^{\mu}}(X) = \sum_{k \geq 0} {\bar r}^{\nu}_{\mu(k)}
[{\bf J},\theta_{0}^{*},q(X)](X)\epsilon^{k} .$$

So that, the analogous expressions will be for the evolution of
the functionals ${\bf U}(X)$ on the ${\cal M}^{\prime}$,
since they are expressed in terms of ${\bf a}(X)$ by the point
substitutions. Besides that, since the system (\ref{qfl})
generates at zero order of $\epsilon$
on the functions from ${\cal M}^{\prime}$ (having the form
(\ref{cz})) the shift of the initial phases $\theta^{\alpha}_{0}$
and does not touch the other parameters, the zero terms do not
present in this series. So that, the Poisson brackets between
$U^{\nu}(X)$ and $J^{\mu}(Y)$ on the ${\cal M}^{\prime}$ will
(as functions of coordinates
${\bf J}(X)$ and $\theta_{0}^{*}(X)$) have the form:
\begin{equation}
\label{dz}
\{U^{\nu}(X),J^{\mu}(Y)\}|_{{\bf G}(\theta,X)=0} =
\sum_{k \geq 1}r^{\nu}_{\mu(k)}[{\bf J},\theta_{0}^{*}] \epsilon^{k}
\end{equation}
(that is $\{U^{\nu}(X),J^{\mu}(Y)\}|_{{\cal M}^{\prime}} = O(\epsilon)$
at $\epsilon \rightarrow 0$ in the coordinates
(${\bf J}(X),\theta_{0}^{*}(X)$).

Lemma 6. \newline
At $Y \neq X_{0}$ we have on the functions from ${\cal M}^{\prime}$,
characterized by the coordinates
${\bf J}(X),\theta_{0}^{*}(X)$, the relations:
\begin{equation}
\label{ez}
\{\theta^{\alpha *}_{0}(X),J^{\nu}(Y)\} = \epsilon \sigma^{\alpha\beta}
[{\bf J},\theta_{0}^{*},\epsilon] ,
\end{equation}
where $\sigma^{\alpha\beta}$ are regular at $\epsilon \rightarrow 0$
functionals of ${\bf J}$ and $\theta_{0}^{*}$.

Proof. \newline
As can be easily seen, the functionals $\theta^{*\alpha}_{0}(X)$ are
expressed, according
to their definition (see (\ref{bc})), by the same (independent
upon $\epsilon$) functions $\tau^{\alpha}(\mbox{$\boldmath \vartheta$}
^{*}(X))$ in terms of the functionals $\vartheta^{*\alpha}(X)$, introduced
by the formula:
$$\vartheta^{*\alpha}(X) = {1 \over (2\pi)^{m}}\int_{0}^{2\pi}\dots\int_
{0}^{2\pi} \sum_{i} \varphi^{i}(\theta,X)\times$$
\begin{equation}
\label{fz}
\times \Phi^{i}_{in \theta^{\alpha}}
(\theta + {1 \over \epsilon}\int_{X_{0}}^
{X}{\bf k}({\bf J}(X^{\prime}))dX^{\prime},
{\bf U}[\mbox{$\boldmath \varphi$}](X)) d^{m}\theta ,
\end{equation}
by which the functionals $\theta^{\alpha}_{0}(X)$
are expressed in terms of
$\mbox{$\boldmath \vartheta$}_{0}(X)$, introduced in (\ref{vth}).
The formulas
\begin{equation}
\label{gz}
\theta^{*\alpha}_{0}(X) =
\tau^{\alpha}(\mbox{$\boldmath \vartheta$}^{*}(X))
\end{equation}
take place in the map on the ${\cal M}^{\prime}$, in which
$\theta_{0}^{*}(X)$ takes at all $\epsilon$
the same (that is independent upon $\epsilon$)
values, that $\theta_{0}(X)$ takes in the map corresponding
to the functions $\tau^{\alpha}$ in the definition of
$\theta^{\alpha}_{0}(X)$ in terms of
$\mbox{$\boldmath \vartheta$}_{0}(X)$.

The evolution of the functionals (\ref{fz}) according to
the flows (\ref{qfl}),
generated by the functionals of type $\int q(X)J^{\mu}(X)dX$,
is determined by the evolution of
$\mbox{$\boldmath \varphi$}(\theta,X)$,
given by the system (\ref{qfl}), and, besides that, by the
evolution of ${\bf J}(X)$ and ${\bf U}[\mbox{$\boldmath \varphi$}](X)$,
which presents in $\mbox{$\boldmath \Phi$}_{in}$ and
in the coordinates ${\bf J}(X)$ and
$\theta_{0}^{*}(X)$ on the ${\cal M}^{\prime}$
is given by the formulas (\ref{jk}) and (\ref{dz}).
After the substitution of
$\mbox{$\boldmath \varphi$}(\theta,X)$
in the form (\ref{cz}) into the system (\ref{qfl}) and integration
with respect to
$\theta$, removing the singularity
at $\epsilon \rightarrow 0$ in the shift of
$\theta$, we can, as in the case with ${\bf U}(X)$, conclude that
the Poisson brackets of $\theta^{*\alpha}(X)$ with $J^{\mu}(Y)$
on the ${\cal M}^{\prime}$ are regular (view (\ref{gz})) at
$\epsilon \rightarrow 0$ functionals of ${\bf J}(X)$ and
$\theta_{0}^{*}(X)$.
That is
$$\{\theta^{*\alpha}_{0}(X),J^{\mu}(Y)\} =
\sum_{k \geq 0}\epsilon^{k} \lambda^{\alpha\mu}
[{\bf J},\theta_{0}^{*}] .$$

Let now $Y \neq X_{0}$. Consider the functionals of type
$\int q(Y^{\prime})J^{\mu}(Y^{\prime})dY^{\prime}$
with the functions $q(Y^{\prime})$,
having the support in the vicinity of $Y$, such that $q(X_{0}) = 0$.
Taking into account the statement (\ref{oxk})
of Lemma 3, it can be easily
seen, that the evolution of the functionals $(\ref{fz})$ according
to the corresponding flows has
on the functions from the ${\cal M}^{\prime}$, characterized by the
coordinates ${\bf J}$ and $\theta_{0}^{*}$,
in the zero
order of $\epsilon$ at $\epsilon \rightarrow 0$ the following form:
$$\vartheta^{*\alpha}_{\tau^{\mu}}(X)|_{\epsilon \rightarrow 0} =
{1 \over (2\pi)^{m}}\int_{0}^{2\pi}\!\!\dots\int_{0}^{2\pi}
[q(X)Q_{\mu}(\mbox{$\boldmath \Phi$}_{in},
k^{\beta}({\bf J})\mbox{$\boldmath \Phi$}_{in\theta^{\beta}},
\dots) \mbox{$\boldmath \Phi$}_{in\theta^{\alpha}}+$$
$$+ \mbox{$\boldmath \Phi$}_{in}
\int_{X_{0}}^{X} \left(\omega^{\beta}_{\mu}({\bf J}(X^{\prime}))
q(X^{\prime})\right)_{X^{\prime}}dX^{\prime} \times
\mbox{$\boldmath \Phi$}_{in\theta^{\alpha}\theta^{\beta}}]
d^{m}\theta =$$
$$= {1 \over (2\pi)^{m}}\int_{0}^{2\pi}\!\!\dots\int_{0}^{2\pi}q(X)
\omega^{\beta}_{\mu}({\bf J}(X))\left[
\mbox{$\boldmath \Phi$}_{in\theta^{\beta}}
\mbox{$\boldmath \Phi$}_{in\theta^{\alpha}} +
\mbox{$\boldmath \Phi$}_{in}
\mbox{$\boldmath \Phi$}_{in\theta^{\alpha}\theta^{\beta}}\right]
d^{m}\theta  = 0 $$
(after the substitution of $\mbox{$\boldmath \varphi$}(\theta,X)$
in form of (\ref{cz}) we use the relation (\ref{fom})
for the functions $Q^{i}_{\mu}$).
The statement of the Lemma immediately follows from this.

For the ${\bf J}(X_{0})$  (at the point $X_{0}$ the depending upon
$\epsilon$ initial phase shift $\theta_{0}$ is absent)
we can obtain by the similar way on the ${\cal M}^{\prime}$
(in the coordinates ${\bf J}(X),\theta_{0}^{*}(X)$):
\begin{equation}
\label{bud}
\{\theta^{*\alpha}(X),J^{\mu}(X_{0})\} = \omega^{\alpha}_
{\mu}({\bf J}(X_{0})) \delta(X-X_{0}) + O(\epsilon) ,
\end{equation}
(since at the zero order of $\epsilon$ the functionals
$\int q(X)J^{\mu}(X)dX$ generate the linear dependence of the initial
phases $\theta_{0}(X)$ upon the time with the frequencies
$q(X)\mbox{$\boldmath \omega$}_{\mu}(X)$ and at the point $X_{0}$:
$\theta^{*}_{0}(X_{0}) \equiv \theta_{0}(X_{0})$).
For the functionals $k^{\alpha}({\bf J}(X))$, view (\ref{rs}),
we have in the coordinates
${\bf J}(X),\theta_{0}^{*}(X)$ on the ${\cal M}^{\prime}$
at all $Y$, including $X_{0}$, the relations:
\begin{equation}
\label{hz}
\{\theta_{0}^{*}(X),k^{\alpha}({\bf J}(Y))\}|_
{{\cal M}^{\prime}} = O(\epsilon) ,\,\,\,\,\,
\epsilon \rightarrow 0 .
\end{equation}

For the functionals $\vartheta^{*\alpha}(X)$, introduced in (\ref{fz}),
we obtain also, using (\ref{az}) and (\ref{hz}), that their Poisson
brackets with each other are regular on ${\cal M}^{\prime}$
in the coordinates ${\bf J}(X),\theta_{0}^{*}(X)$
at $\epsilon \rightarrow 0$, and, so that, the same property is valid
for the brackets of type:
$\{\theta^{*\alpha}_{0}(X),\theta^{\beta*}_{0}(Y)\}$,
that is
\begin{equation}
\label{iz}
\{\theta^{*\alpha}_{0}(X),\theta^{\beta*}_{0}(Y)\}|_{{\cal M}^{\prime}} =
\gamma^{\alpha\beta}[{\bf J},\theta_{0}^{*},\epsilon](X,Y) ,
\end{equation}
where $\gamma^{\alpha\beta}(X,Y)$ are regular at $\epsilon
\rightarrow 0$ functional of ${\bf J}$ и $\theta_{0}^{*}$.
We shall not need for the more precise information
about the brackets of this kind.

Let now consider the Dirac procedure of restriction of the
bracket (\ref{esk}) on the submanifold ${\cal M}^{\prime}$,
using the coordinates ${\bf J}(X),\theta_{0}^{*}(X)$ and
$G^{i}(\theta,X)$ in $\Delta_{\delta}$.

For the Dirac restriction of the bracket
(\ref{esk}) on ${\cal M}^{\prime}$
we must find for the functionals $J^{\nu}(X)$ and
$\theta^{*\alpha}_{0}(X)$ the additions of type:
\begin{equation}
\label{jz}
V^{\nu}(X) = {1 \over (2\pi)^{m}}\int_{0}^{2\pi}\!\!\dots
\int_{0}^{2\pi} \int v^{\nu}_{i}(X,Y,\theta,\epsilon)
G^{i}(\theta,Y) d^{m}\theta dY ,
\end{equation}
\begin{equation}
\label{jjzz}
W^{\alpha}(X) =
{1 \over (2\pi)^{m}}\int_{0}^{2\pi}\!\!\dots\int_{0}^{2\pi} \int
w^{\alpha}_{i}(X,Y,\theta,\epsilon)
G^{i}(\theta,Y) d^{m}\theta dY ,
\end{equation}
such that the flows, generated by the functionals ${\tilde J}^{\nu}(X) =
J^{\nu}(X) + V^{\nu}(X)$ and ${\tilde \theta}^{\alpha}_{0}(X) =
\theta^{*\alpha}_{0}(X) + W^{\alpha}(X)$, leave ${\cal M}^{\prime}$
invariant. After that we must put on ${\cal M}^{\prime}$:
\begin{equation}
\label{kz}
\{J^{\nu}(X),J^{\mu}(Y)\}^{*} = \{{\tilde J}^{\nu}(X),
{\tilde J}^{\mu}(Y)\}|_{{\cal M}^{\prime}}({\bf J},\theta_{0}^{*}) ,
\end{equation}
\begin{equation}
\label{kzz}
\{\theta^{*\alpha}_{0}(X),\theta^{\beta*}(Y)\}^{*} =
\{{\tilde \theta}^{\alpha}_{0}(X),{\tilde \theta}^{\beta}_{0}(Y)\}
|_{{\cal M}^{\prime}}({\bf J},\theta_{0}^{*}) ,
\end{equation}
\begin{equation}
\label{kzzz}
\{J^{\nu}(X),\theta^{*\alpha}_{0}(Y)\}^{*} =
\{{\tilde J}^{\nu}(X),{\tilde \theta}^{*\alpha}_{0}(Y)\}|_
{{\cal M}^{\prime}}({\bf J},\theta_{0}^{*}) .
\end{equation}

On the functions $v^{\nu}_{i}(X,Y,\theta,\epsilon)$
and $w^{\alpha}_{i}(X,Y,\theta,\epsilon)$
we obtain, respectively, the relations:
$${1 \over (2\pi)^{m}}\int_{0}^{2\pi}\!\!\dots\int_{0}^{2\pi} \int
v^{\nu}_{j}(Y,Z,\theta^{\prime},\epsilon) \times$$
$$\times \{G^{i}({\bf U}[\mbox{$\boldmath \varphi$}](X),
\mbox{$\boldmath \varphi$}(\theta,X),\dots),
G^{j}({\bf U}[\mbox{$\boldmath \varphi$}](Z),
\mbox{$\boldmath \varphi$}(\theta^{\prime},Z),\dots)\}
d^{m}\theta^{\prime}dZ |_{{\cal M}^{\prime}} =$$
\begin{equation}
\label{mz}
= - \{G^{i}({\bf U}[\mbox{$\boldmath \varphi$}](X),
\mbox{$\boldmath \varphi$}(\theta,X),
\dots),J^{\nu}[\mbox{$\boldmath \varphi$}](Y)\}|_{{\cal M}^{\prime}}
\end{equation}
$${1 \over (2\pi)^{m}}\int_{0}^{2\pi}\!\!\dots\int_{0}^{2\pi} \int
w^{\alpha}_{j}(Y,Z,\theta^{\prime},\epsilon) \times$$
$$\times \{G^{i}({\bf U}[\mbox{$\boldmath \varphi$}](X),
\mbox{$\boldmath \varphi$}(\theta,X),
\dots),G^{j}({\bf U}[\mbox{$\boldmath \varphi$}])(Z),
\mbox{$\boldmath \varphi$}(\theta^{\prime},Z),
\dots)\} d^{m}\theta^{\prime}dZ |_
{{\cal M}^{\prime}} =$$
\begin{equation}
\label{mzz}
= - \{G^{i}({\bf U}[\mbox{$\boldmath \varphi$}](X),
\mbox{$\boldmath \varphi$}(\theta,X),
\dots),\theta^{*\alpha}_{0}[\mbox{$\boldmath \varphi$}](Y)\}|_
{{\cal M}^{\prime}}
\end{equation}

In the calculation of the Poisson bracket
of constraints $G^{i}(\theta,X)$
with any functional on ${\cal M}^{\prime}$ we can use the property,
that on ${\cal M}^{\prime}$ the values standing in the brackets
in the definition of constraints $G^{i}(\theta,X)$
in terms of $\mbox{$\boldmath \varphi$}(X)$
(see (\ref{ccc})) become zeros and, so that, in the calculation of
the brackets of type
$\{G^{i}(\theta,X),C(\theta^
{\prime},Z)\}$ on ${\cal M}^{\prime}$ we can omit the brackets
of the kernels of the operators ${\hat {\bf L}}$,
presenting in (\ref{ccc}), with $C(\theta^{\prime},Z)$
and replace the kernels of ${\hat {\bf L}}$ from the Poisson brackets
in the form of multipliers according to the Leibnitz identity.

Regarding also the relations (\ref{hz}), (\ref{az}), and
(\ref{jj}), (\ref{ez}), (\ref{bud}), (\ref{iz}) for the functionals
${\bf J}$ and $\theta_{0}^{*}$,
presenting in (\ref{ccc}), we can see that the Poisson brackets
of type $\{G^{i}(\theta,X), G^{j}(\theta^{\prime},Z)\}$,
$\{G^{i}(\theta,X),J^{\nu}(Y)\}$ and
$\{G^{i}(\theta,X),\theta^{*\alpha}_{0}(Y)\}$
can on the submanifold ${\cal M}^{\prime}$ with the coordinates
${\bf J}(X),\theta_{0}^{*}(X)$
be represented in the most general form at
$\epsilon \rightarrow 0$ by the asymptotic series of type:
$$\{G^{i}(\theta,X),
G^{j}(\theta^{\prime},Z)\}|_{{\cal M}^{\prime}} =$$
$$= {1 \over (2\pi)^{2m}} \int_{0}^{2\pi}\!\!\!\!\dots\int_{0}^{2\pi}
{\tilde L}^{i}_{s [{\bf U}[{\bf J},\theta_{0}^{*},\epsilon]]}
\left(\theta + \theta_{0}^{*}(X) +
{{\bf s}(X) \over \epsilon},\tau + \theta_{0}^{*}(X) +
{{\bf s}(X) \over \epsilon},X\right)\times$$
$$\times \left(\sum_{{\bf n},k\geq p\geq 0} M^{sl}_{(k),(p),{\bf n}}
(\tau + \theta^{*}_{0}(X) + {{\bf s}(X) \over \epsilon},
[{\bf J},\theta^{*}_{0}]) \epsilon^{k} \delta^{(p)}(X-Z)
\delta_{{\bf n}\theta}(\mbox{$\boldmath \tau$} -
\mbox{$\boldmath \zeta$})\right)\times$$
\begin{equation}
\label{nz}
\times {\tilde L}^{i}_{l [{\bf U}[{\bf J},\theta_{0}^{*},\epsilon]]}
\left(\theta^{\prime} + \theta_{0}^{*}(Z) + {{\bf s}(Z) \over \epsilon},
\mbox{$\boldmath \zeta$} +
\theta_{0}^{*}(Z) + {{\bf s}(Z) \over \epsilon},Z\right)
d^{m}\mbox{$\boldmath \tau$} d^{m}\mbox{$\boldmath \zeta$} ,
\end{equation}
${\bf s}(X)$ and ${\bf s}(Z)$ denote the integrals
$\int_{X_{0}}^{X}{\bf k}({\bf J}(X^{\prime}))dX^{\prime}$
and $\int_{X_{0}}^{Z}{\bf k}({\bf J}(X^{\prime}))dX^{\prime}$, 
and also:
$$\{G^{i}(\theta,X),{\bf J}^{\nu}(Y)\}|_
{{\cal M}^{\prime}} =$$
$$= {1 \over (2\pi)^{m}}\int_{0}^{2\pi}\!\!\!\!\dots\int_{0}^{2\pi}
{\tilde L}^{i}_{s [{\bf U}[{\bf J},\theta_{0}^{*},\epsilon]]}
\left(\theta + \theta_{0}^{*}(X) +
{{\bf s}(X) \over \epsilon},\mbox{$\boldmath \tau$} + \theta_{0}^{*}(X) +
{{\bf s}(X) \over \epsilon},X\right)\times$$
\begin{equation}
\label{oz}
\times\left(\sum_{k\geq p\geq0}S^{s\nu}_{(k),(p)}(\tau + \theta^{*}_{0}(X)
+ {{\bf s}(X) \over \epsilon},[{\bf J},\theta^{*}_{0}])
\epsilon^{k} \delta^{(p)}(X-Y)\right) d^{m}\mbox{$\boldmath \tau$} ,
\end{equation}
$$\{G^{i}(\theta,X),\theta^{*\alpha}(Y)\}|_{{\cal M}^{\prime}} =$$
$$= {1 \over (2\pi)^{m}}\int_{0}^{2\pi}\!\!\!\!\dots\int_{0}^{2\pi}
{\tilde L}^{i}_{s [{\bf U}[{\bf J},\theta_{0}^{*},\epsilon]]}
\left(\theta + \theta_{0}^{*}(X) + {{\bf s}(X) \over \epsilon},
\mbox{$\boldmath \tau$} + \theta_{0}^{*}(X) +
{{\bf s}(X) \over \epsilon},X\right)\times $$
\begin{equation}
\label{pz}
\times \left(\sum_{k\geq p\geq 0}T^{s\alpha}_{(k),(p)}
(\tau + \theta^{*}_{0}(X) + {{\bf s}(X) \over \epsilon},
[{\bf J},\theta^{*}_{0}])
\epsilon^{k} \delta^{(p)}(X-Y)\right) d^{m}\mbox{$\boldmath \tau$} .
\end{equation}
(All the functions in the sums within brackets,
depending upon $Z$ and $Y$, are replaced by the
functions, depending upon $X$, according to the formulas of type:
$\epsilon\delta^{\prime}(X-Z)f(Z) =
\epsilon f(X)\delta^{\prime}(X-Z) + \epsilon f_{X}(X)\delta(X-Z)$.
As can be easily seen, all differentiations with respect to $X$
appear with the multiplier $\epsilon$.)

The functions
\begin{equation}
\label{qk}
{\tilde {\mbox{$\boldmath \kappa$}}}^{q}_{[{\bf U}[{\bf J},
\theta_{0}^{*},\epsilon]]}
(\theta + \theta_{0}^{*}(X) +
{1 \over \epsilon}\int_{X_{0}}^{X}{\bf k}({\bf J}
(X^{\prime}))dX^{\prime},X)
\end{equation}
and
\begin{equation}
\label{kq}
{\tilde {\mbox{$\boldmath \kappa$}}}^{q}_
{[{\bf U}[{\bf J},\theta_{0}^{*},\epsilon]]}
(\theta +\theta_{0}^{*}(Z) +
{1 \over \epsilon}\int_{X_{0}}^{Z}{\bf k}({\bf J}
(X^{\prime})dX^{\prime},Z)
\end{equation}
are respectively left and right eigen
vectors on the ${\cal M}^{\prime}$
of the linear operator in the space of
$2\pi- $ periodic with respect to $\theta$ functions with the kernel
$\{G^{i}(\theta,X),G^{j}(\theta^{\prime},Z)\}$
and correspond to zero eigen values. The functions
$\mbox{$\boldmath \kappa$}_{[{\bf U}[{\bf J},\theta_
{0}^{*},\epsilon]]}
(\theta + \theta_{0}^{*}(X) +
{1 \over \epsilon}\int_{X_{0}}^{X}{\bf k}({\bf J}(X^{\prime}))dX^
{\prime},X)$,
as can be easily seen from (\ref{oz}) and (\ref{pz}), are also
orthogonal (at all $\epsilon$) to the right
parts of (\ref{mz}),(\ref{mzz}),
and,so that, the systems (\ref{mz}),(\ref{mzz})
are resolvable in the generic case.
The solutions $v^{\nu}_{j}(X,Y,\theta)$
and $w^{\alpha}_{j}(X,Y,\theta)$
are defined modulo the arbitrary linear combination of vectors
$\mbox{$\boldmath \kappa$}_{[{\bf U}[{\bf J},\theta_{0}^{*},\epsilon]]}
(\theta + \theta_{0}^{*}(Y) + {1 \over \epsilon}\int_{X_{0}}^
{Y}{\bf k}({\bf J}(X^{\prime})dX^{\prime},Y)$,
but it does not influence, view (\ref{rassl}), neither on the form
of the additions
$V^{\nu}(X)$ and $W^{\alpha}(X)$, nor on the Dirac restriction
of the bracket (\ref{esk}) on ${\cal M}^{\prime}$ according to the
formulas (\ref{kz})-(\ref{kzzz}), since in the calculation of the
brackets of $V^{\nu}(X)$ and
$W^{\alpha}(X)$ with any functionals on the ${\cal M}^{\prime}$
the contraction of this linear combination
with the kernel of corresponding
$L^{i}_{s [{\bf U}[{\bf J},\theta_{0}^{*},
\epsilon]]}(\dots)$
will be replaced (according to the said above)
from the Poisson bracket as
a multiplier which is equal to zero on ${\cal M}^{\prime}$.

For the unique determination of the functions:
\begin{equation}
\label{vv}
v^{\nu}_{j}(X,Y,\theta,\epsilon) =
{\bar v}^{\nu}_{j}(X,Y,\theta +
\theta_{0}^{*}(Y) +
{1 \over \epsilon}\int_{X_{0}}^{Y}{\bf k}({\bf J}(X^{\prime}))dX^
{\prime},\epsilon)
\end{equation}
and
\begin{equation}
\label{ww}
w^{\alpha}_{j}(X,Y,\theta,\epsilon) =
{\bar w}^{\alpha}_{j}(X,Y,\theta +
\theta_{0}^{*}(Y) +
{1 \over \epsilon}\int_{X_{0}}^{Y}{\bf k}({\bf J}(X^{\prime}))dX^
{\prime},\epsilon)
\end{equation}
in the coordinates ${\bf J}(X)$ and $\theta^{*}_{0}(X)$ on
${\cal M}^{\prime}$
we put $N+m$ additional relations on them, demanding from them to be
also orthogonal at all $X$ and $Y$ to the corresponding functions:
\begin{equation}
\label{kk}
{\tilde {\mbox{$\boldmath \kappa$}}}^{q}_
{[{\bf U}[{\bf J},\theta_
{0}^{*},\epsilon]]}
(\theta + \theta_{0}^{*}(Y) +
{1 \over \epsilon}\int_{X_{0}}^{Y}{\bf k}({\bf J}(X^{\prime}))dX^
{\prime},Y,\epsilon) ,
\end{equation}
that is
\begin{equation}
\label{abc}
{1 \over (2\pi)^{m}}\int_{0}^{2\pi}\!\!\dots\int_{0}^{2\pi}
\sum_{j}{\bar v}^{\nu}_{j}(X,Y,\theta,\epsilon)
{\tilde {\mbox{$\boldmath \kappa$}}}^{q}_
{j [{\bf U}[{\bf J},\theta_{0}^{*},\epsilon]]}
(\theta,Y,\epsilon) d^{m}\theta = 0 ,
\end{equation}
\begin{equation}
\label{bcd}
{1 \over (2\pi)^{m}}\int_{0}^{2\pi}\!\!\dots\int_{0}^{2\pi}
\sum_{j}{\bar w}^{\alpha}_{j}(X,Y,\theta,\epsilon)
{\tilde {\mbox{$\boldmath \kappa$}}}^{q}_
{j [{\bf U}[{\bf J},\theta_{0}^{*},\epsilon]]}
(\theta,Y,\epsilon) d^{m}\theta = 0 ,
\end{equation}
at all $X,Y,\epsilon$ on ${\cal M}^{\prime}$.

After the substitution of values $v^{\nu}_{j}(X,Y,\theta,\epsilon)$
and $w^{\alpha}_{j}(X,Y,\theta,\epsilon)$
in the form (\ref{vv}),(\ref{ww}) into the systems (\ref{mz}),
(\ref{mzz}), where
$\{G^{i}(\theta,X),G^{j}(\theta^{\prime},Z)\}$,\linebreak
$\{G^{i}(\theta,X),J^{\nu}(Y)\}$ and
$\{G^{i}(\theta,X),\theta^{*\alpha}_{0}(Y)\}$
are taken in the form (\ref{nz}), (\ref{oz}) and (\ref{pz})
respectively, on the functions ${\bar v}^{\nu}_{j}$
and ${\bar w}^{\alpha}_{j}$ (after all differentiations with respect to
$X$ in (\ref{mz}),(\ref{mzz}), appearing in every case with the
multiplier $\epsilon$,
the singular at $\epsilon \rightarrow 0$ phase shift
$\theta_{0}^{*}(X) +
{1 \over \epsilon}\int_{X_{0}}^{X}{\bf k}({\bf J}
(X^{\prime}))dX^{\prime}$,
which presents in all functions, depending upon $\theta$,
can be omitted) we obtain the linear nonhomogeneous systems,
which can be represented in form of regular at
$\epsilon \rightarrow 0$ asymptotic series with respect to $\epsilon$.
As was mentioned above, in the generic case corresponding to the
nonsingularity of the matrix of Poisson brackets of constraints
in the finite-dimensional case, these systems are resolvable
at all $\epsilon$,
and, in the presence of the additional relations (\ref{abc}),(\ref{bcd}),
the functions ${\bar v}(X,Y,\theta,\epsilon)$ and
${\bar w}(X,Y,\theta,\epsilon)$ can be uniquely determined
on the submanifold ${\cal M}^{\prime}$ with the coordinates
${\bf J}(X),\theta_{0}^{*}(X)$
(the systems (\ref{mz}),(\ref{mzz}) depend of them as of parameters)
in the form of regular at $\epsilon \rightarrow 0$ asymptotic series
with respect to $\epsilon$, that is
\begin{equation}
\label{cde}
{\bar v}^{\nu}_{j}(X,Y,\theta,[{\bf J},
\theta_{0}^{*}],\epsilon)
= \sum_{k\geq 0}{\bar v}^{\nu}_{j(k)}
(X,Y,\theta,[{\bf J},\theta_{0}^{*}]) \epsilon^{k} ,
\end{equation}
\begin{equation}
\label{efg}
{\bar w}^{\alpha}_{j}(X,Y,\theta,
[{\bf J},\theta_{0}^{*}],\epsilon) =
\sum_{k\geq 0}{\bar w}^{\alpha}_{j(k)}
(X,Y,\theta,[{\bf J},\theta_{0}^{*}])
\epsilon^{k} .
\end{equation}

Remark. \newline
Very frequently the Poisson brackets (\ref{bracket}) for the initial
system (\ref{system}), such as the Gardner - Zakharov - Faddeev bracket:
$$\{\varphi(x),\varphi(y)\} = \delta^{\prime}(x-y) ,$$
or Magri - Lenard bracket:
$$\{\varphi(x),\varphi(y)\}_{ML} = -{1 \over 2}\delta^
{\prime\prime\prime}(x-y) + (\varphi(x) + \varphi(y)) \delta^
{\prime}(x-y) ,$$
have such a form, that the corresponding to them bracket (\ref{esk})
is degenerated at zero order of $\epsilon$
in the coordinates $\mbox{$\boldmath \varphi$}(\theta,X)$.
This degeneracy arises because of the presence
of the derivatives of $\delta -$ functions with respect to $x$
(that is the operators $\partial/\partial x$) or the derivatives with
respect to $x$ of functions
$\mbox{$\boldmath \varphi$}(x)$ in every term of such brackets, which
arise in (\ref{esk}) with the multipliers $\epsilon$.
However, in the introduced above coordinates
$J^{\nu}(X), \theta^{*\alpha}_{0}(X)$ and
$G^{i}(\theta,X)$ in the vicinity of
${\cal M}^{\prime}$ the operators $\epsilon\partial/\partial X$,
being applied to the functions
(\ref{rs}) on the ${\cal M}^{\prime}$, contain the nonvanishing at
$\epsilon \rightarrow 0$ value of type $k^{\alpha}({\bf J}(X)){\partial
\over \partial \theta^{\alpha}}$, which permits to assume in 
the generic
case (for the generic constraints) that the systems
on the functions ${\bar v}^{\nu}_{j}$ and ${\bar w}^{\alpha}_{j}$ are
nondegenerate in the zero order of
$\epsilon$, and to write for the functions ${\bar v}^{\nu}_{j}$
and ${\bar w}^{\alpha}_{j}$ the regular at $\epsilon \rightarrow 0$
asymptotic expansions (\ref{cde}) and (\ref{efg}).

More precisely, the expansions (\ref{cde}) and (\ref{efg}) have the form:
\begin{equation}
\label{fgh}
{\bar v}^{\nu}_{j}(X,Y,\theta,\epsilon) =
\sum_{k\geq p\geq 0}{\bar v}^{\nu}_{j(k),(p)}(X,\theta,
[{\bf J},\theta_{0}^{*}]) \epsilon^{k}
\delta^{(p)}(X-Y) ,
\end{equation}
\begin{equation}
\label{ghi}
{\bar w}^{\alpha}_{j}(X,Y,\theta,\epsilon) =
\sum_{k\geq p\geq 0}{\bar w}^{\alpha}_{j(k),(p)}
(X,\theta,[{\bf J},\theta_{0}^{*}])
\epsilon^{k} \delta^{(p)}(X-Y)
\end{equation}
(here at every given $k$ $p$ runs the finite number of values
$p\leq k$), the values $V^{\nu}(X)$ and $W^{\alpha}(X)$, as can be
easily seen, do not contain the generalized functions. The substitution
of (\ref{fgh}) and (\ref{ghi}) in the corresponding systems
(\ref{mz}),(\ref{mzz}) gives in the
k-th order of $\epsilon$ and at any given
$X$ and $p$ (the coefficients with the derivatives of
$\delta -$ functions  must
be equal to each other similar to the coefficients of powers of
$\epsilon$) the linear (differential with respect to
$\theta$) nonhomogeneous systems on the functions
${\bar v}^{\nu}_{j(k),(p)}(X,\theta)$ and
${\bar w}^{\alpha}_{j(k),(p)}(X,\theta)$,
depending of ${\bf J}$ and $\theta_{0}^{*}$
as of the parameters. The right sides of these systems depend
(in the linear manner) upon the previous
${\bar {\bf v}}_{(k^{\prime}),(q)}$
and ${\bar {\bf w}}_{(k^{\prime}),(q)}$ с $k^{\prime}<k$.
Under the assumption made above about the unique resolvability
of the systems (\ref{mz}),(\ref{mzz}) at all $\epsilon$
in the presence of the additional conditions
(\ref{abc}) and (\ref{bcd}), corresponding to the nonsingularity
of the matrix of Poisson brackets of constraints on the
${\cal M}^{\prime}$,
the system described now will be uniquely resolvable
under the additional conditions (\ref{abc}) and (\ref{bcd})
in the corresponding order of $\epsilon$,
and,so that, the series (\ref{fgh}), (\ref{ghi}) may be constructed
by successive approximations.

Besides that, as was mentioned above, the flows (\ref{qfl}), generated
by the functionals $\int q(X)J^{\mu}(X)dX$ on the functions
of form (\ref{cz}), leave invariant the submanifold ${\cal M}^{\prime}$
at zero order of $\epsilon$,
generating on it the linear dependence
of the initial phases upon the times
$\tau^{\mu}$, and, so that, in the
expression (\ref{oz}) and, respectively,
in the right side of the system
(\ref{mz}) the zero term of $\epsilon$ is absent
$(S^{s\nu}_{0} \equiv 0)$. From this we
may conclude that, under the
assumption of the unique resolvability
of the systems on the functions
${\bar v}^{\nu}_{j}$ at zero order of $\epsilon$
(in the coordinates ${\bf J},\theta_{0}^{*}$ on
${\cal M}^{\prime}$): ${\bar v}^{\nu}_{j(0),(p)} \equiv 0$, that is
for the functions $v^{\nu}_{j}$ and $w^{\alpha}_{j}$ we may write:
$$v^{\nu}_{j}(X,Y,\theta,[{\bf J},
\theta_{0}^{*}],\epsilon) =$$
\begin{equation}
\label{prs}
= \sum_{k\geq p\geq 1} {\bar v}^{\nu}_{j(k),(p)}
(X,\theta + \theta_{0}^{*}(Y) +
{1 \over \epsilon}\int_{X_{0}}^{Y}{\bf k}({\bf J}(X^{\prime}))dX^{\prime},
[{\bf J},\theta_{0}^{*}]) \epsilon^{k} \delta^{(p)}(X-Y) ,
\end{equation}
$$w^{\alpha}_{j}(X,Y,\theta,[{\bf J},
\theta_{0}^{*}],\epsilon) =$$
\begin{equation}
\label{rst}
= \sum_{k\geq p\geq 0} {\bar w}^{\alpha}_{j(k),(p)}
(X,\theta + \theta_{0}^{*}(Y) +
{1 \over \epsilon}\int_{X_{0}}^{Y}{\bf k}({\bf J}(x^{\prime}))dX^{\prime},
[{\bf J},\theta_{0}^{*}]) \epsilon^{k} \delta^{(p)}(X-Y) .
\end{equation}
The more precise information about the functions $v^{\nu}_{j}$ and
$w^{\alpha}_{j}$ will be not necessary for us.

Let us now formulate the main result of this paper.

Theorem 1. \newline
Suppose that for the system (\ref{system}), Hamiltonian with respect
to bracket (\ref{bracket}) and having the family of m - phase solutions
and the sufficient number of conservation laws (\ref{laws}),
take place all the described above properties of generic situation,
concerning the functional independence of the parameters $U^{\nu}$
on this family and the possibility of the expression of
parameters ${\bf k},\mbox{$\boldmath \omega$}$ and ${\bf r}$
in terms of them, and, besides that,
takes place the relation, connecting the
number of the annihilators of bracket
(\ref{bracket}) with the number of introduced above additional
parameters $r^{1},\dots,r^{g}$ $(g=p)$.
Then, under the assumption of regularity ((A) and (B))
of the introduced previously submanifolds
${\cal M}^{\nu}_{\mbox{$\boldmath \omega$}^{\nu},{\bf k}}$,
(that is the possibility of constructing for any system (\ref{qfl}),
generated by the functional $\int q(X)J^{\nu}(X)dX$, of the asymptotical
solutions (\ref{kl})), and the analogous regularity of
${\cal M}$ with the nonsingularity of matrix of Poisson brackets
of constraints $\{G^{i}(\theta,X),G^{j}(\theta^{\prime},Z)\}$
on ${\cal M}^{\prime}$ at zero order of $\epsilon$ in the coordinates
${\bf J},\theta_{0}^{*}$ (that is the possibility of constructing
of functions $v^{\nu}_{j}$
and $w^{\alpha}_{j}$ in form of the asymptotical series
(\ref{prs}),(\ref{rst})):
\newline 1) The Dubrovin - Novikov bracket, defined by the formula
(\ref{sdn}):
$$\{U^{\nu}(X),U^{\mu}(Y)\} = \langle A^{\nu\mu}_{1}\rangle({\bf U}(X))
\delta^{\prime}(X-Y) + {\partial\langle Q^{\nu\mu}\rangle \over
\partial U^{\lambda}}({\bf U}(X)) U^{\lambda}_{X} \delta(X-Y) ,$$
satisfies to the Jacobi identity. \newline
2) For the bracket (\ref{sdn}) take place the following relations:
\begin{equation}
\label{stu}
\{k^{\alpha}({\bf U}(X)),k^{\beta}({\bf U}(Y))\} = 0 , \,\,\,
\{k^{\alpha}({\bf U}(X)),n_{q}({\bf U}(Y))\} = 0 ,
\end{equation}
\begin{equation}
\label{sstu}
\{k^{\alpha}({\bf U}(X)),U^{\mu}(Y))\} = \omega^{\alpha}_{\nu}({\bf U}(X))
\delta^{\prime}(X-Y) + \omega^{\alpha}_{\nu X}({\bf U}(X)) \delta(X-Y) .
\end{equation}

Proof. \newline
If the conditions, formulated above,
are valid, we can restrict the Poisson
bracket (\ref{esk}) on the submanifold ${\cal M}^{\prime}$ with the
coordinates ${\bf J}(X),\theta_{0}^{*}(X)$ according to the described
procedure. According to the formula (\ref{ogran}), for the restricted
on the ${\cal M}^{\prime}$ bracket we shall have the following formulas:
$$\{J^{\nu}(X),J^{\mu}(Y)\}^{*} = \{J^{\nu}(X),J^{\mu}(Y)\}|_
{{\cal M}^{\prime}}({\bf J},\theta_{0}^{*}) -$$
$$- {1 \over (2\pi)^{2m}}\int_{0}^{2\pi}\!\!\dots\int_{0}^{2\pi}\int\int
v^{\nu}_{i}(X,{\bar X},\theta,\epsilon)\times
\{G^{i}(\theta,{\bar X}),
G^{j}(\theta,{\bar Y})\}\times$$
$$\times v^{\mu}_{j}(Y,{\bar Y},\theta^{\prime},\epsilon)
d{\bar X} d{\bar Y}
d^{m}\theta d^{m}\theta^{\prime} ,$$
and the analogous formulas for
$\{J^{\nu}(X),\theta^{*\alpha}_{0}(Y)\}^{*}$ and
$\{\theta^{*\alpha}_{0}(X),\theta^{*\beta}_{0}(Y)\}^{*}$
(with the replacing of functions $v^{\nu}_{i}, v^{\mu}_{j}$ to
$w^{\alpha}_{i}, w^{\beta}_{j}$ for the functions
$\theta^{*\alpha}_{0}(X)$ and $\theta^{*\beta}_{0}(Y)$).
As previously, it can be easily seen that in the integration
of functions (\ref{prs}), (\ref{rst}) and (\ref{nz})
the singular at $\epsilon \rightarrow 0$ phase shift
$\theta_{0}^{*}(X) + {1 \over \epsilon}\int_{X_{0}}^
{X}{\bf k}({\bf J}(X^{\prime}))dX^{\prime}$,
which presents in all functions, depending upon
$\theta$ and $\theta^{\prime}$
(after the integration with respect to $d{\bar X}$ and $d{\bar Y}$),
is unessential, and the Poisson bracket $\{\dots,\dots\}^{*}$
on ${\cal M}^{\prime}$ is regular at $\epsilon \rightarrow 0$
in the coordinates ${\bf J}(X)$ and $\theta_{0}^{*}(X)$.
Using the relations (\ref{jj}), (\ref{ez}), (\ref{iz}),
and (\ref{prs}), (\ref{rst}), we obtain for the bracket
$\{\dots,\dots\}^{*}$ in the coordinates
${\bf J},\theta_{0}^{*}$:
$$\{J^{\nu}(X),J^{\mu}(Y)\}^{*} =$$
\begin{equation}
\label{qdq}
= \epsilon \left(\langle A^{\nu\mu}_
{1}\rangle({\bf J}(X)) \delta^{\prime}(X-Y) +
({\partial \over \partial X}
\langle Q^{\nu\mu}\rangle({\bf J}(X))) \delta(X-Y)\right)
+ O(\epsilon^{2}),
\end{equation}
\begin{equation}
\label{dqd}
\{J^{\nu}(X),\theta^{*\alpha}_{0}(Y)\}^{*} = O(\epsilon)
\end{equation}
at $X \neq X_{0}$,
\begin{equation}
\label{qfq}
\{\theta^{*\alpha}_{0}(X),\theta^{*\beta}_{0}(Y)\}^{*} = O(1)
\end{equation}
at $\epsilon \rightarrow 0$.

>From this it can be seen that the Jacobi identities, containing only
the functionals
${\bf J}$ at the points $X,Y,Z$, which do not coincide with $X_{0}$,
coincide in the first nonvanishing order of $\epsilon$
(at $\epsilon^{2}$) with the corresponding Jacobi identities
for the bracket (\ref{sdn}) on the space of fields $U^{\nu}(X)$.
Regarding now that the point
$X_{0}$ is arbitrary, and the expression (\ref{sdn})
does not depend upon $X_{0}$, we obtain that the bracket (\ref{sdn})
satisfies the Jacobi identity.
Its skew-symmetry is a trivial corollary from the skew-symmetry of
the bracket (\ref{bracket}).

The relations (\ref{stu}),(\ref{sstu}) now follow from
(\ref{az}), (\ref{bz}) and (\ref{kjk}) respectively.

Theorem is proved.

Theorem 2. \newline
Suppose now that under the assumptions formulated in Theorem 1
we have two different sets
$\{I^{1},\dots,I^{N}\}, \{{\bar I}^{1},\dots,{\bar I}^{N}\}$
of $N$ conservation laws of system (\ref{system}) of form (\ref{laws}),
(some of the integrals of these two sets may coincide with each other).
Then, the Dubrovin - Novikov brackets obtained with the aid
of these two sets are coincide.
That is, if $U^{\nu} = \langle{\cal P}^{\nu}\rangle ,
{\bar U}^{\nu} = \langle{\bar {\cal P}}^{\nu}\rangle$, where
${\cal P}^{\nu}, {\bar {\cal P}}^{\nu}$ - are
the densities of the integrals
of the first and second sets respectively, $\langle\dots\rangle$
is the averaging on the family of m - phase solutions of (\ref{system}),
then the transformation of bracket (\ref{sdn}), obtained with the aid
of the set $\{I^{1},\dots,I^{N}\}$, to the coordinates ${\bar U^{\nu}(X)}$
on the ${\cal M}$, expressed by the point substitutions:
\begin{equation}
\label{fqf}
{\bar U}^{\nu} = {\bar u}^{\nu}({\bf U})
\end{equation}
in terms of ${\bf U}(X)$, gives the
Dubrovin - Novikov bracket, obtained with
the aid of the set $\{{\bar I}^{1},\dots,{\bar I}^{N}\}$.

Proof.\newline
In the case of the generic situation under consideration,
the Dirac restriction of the bracket (\ref{esk}) on the submanifold
${\cal M}^{\prime}$
is uniquely determined, that is, the brackets (\ref{qdq}) - (\ref{qfq}),
obtained in the coordinates $({\bf J},\theta^{*}_{0})$ and
$({\bar {\bf J}},\theta^{*}_{0})$ (corresponding to the first and
second sets of integrals respectively) in the vicinity of
${\cal M}^{\prime}$, transform into each other under the corresponding
transformations of the coordinates
друга координат $({\bf J},\theta^{*}_{0})$ and
$({\bar {\bf J}},\theta^{*}_{0})$ on the ${\cal M}^{\prime}$.
Regarding (\ref{ef}) and the similar relation for
${\bar {\bf U}}$ and $({\bar {\bf J}},\theta^{*}_{0})$,
we can conclude that the transition from the coordinates ${\bf J}(X)$
to ${\bar {\bf J}}(X)$ has the form:
$${\bar J}^{\nu}(X) = {\bar u}^{\nu}({\bf J}(X)) + \sum_{k\geq1}
\epsilon^{k} {\bar J}^{\nu}_{(k)}({\bf J},{\bf J}_{X},\dots,{\bf J}_{kX},
\theta^{*}_{0X},\dots,\theta^{*}_{0kX}) ,$$
and, so that, the transition from the brackets (\ref{qdq}) to
$\{{\bar J}^{\nu}(X),{\bar J}^{\mu}(Y)\}$ coincides
for such transformation (if $X,Y \neq X_{0}$) in the first nonvanishing
order of $\epsilon$ (at zero power of $\epsilon$) with the
corresponding transformation of Dubrovin - Novikov bracket under the
substitution (\ref{fqf}), which proves the Theorem.

The evolution of densities ${\cal P}^{\nu}(x)$ according to the
flows (\ref{fluxes}), generated by the integrals $I^{\mu}$,
as can be easily seen from (\ref{skob}), has the form:
\begin{equation}
\label{qfqf}
{\cal P}^{\nu}_{\tau^{\mu}}(\mbox{$\boldmath \varphi$},
\mbox{$\boldmath \varphi$}_{x},\dots) = \partial_{x}
Q^{\nu\mu}(\mbox{$\boldmath \varphi$},\mbox{$\boldmath \varphi$}_{x},
\dots) .
\end{equation}
The flows, generated by the functionals $\int U^{\mu}(X)dX$
on the space of fields ${\bf U}(X)$ with the aid of Dubrovin -
Novikov bracket, has the form:
\begin{equation}
\label{gqg}
U^{\nu}_{T^{\mu}} = \partial_{X} \langle Q^{\nu\mu}\rangle({\bf U}) ,
\end{equation}
and, by such a way, represent (see (\ref{up})) the Whitham's
equations for m - phase solutions of the Hamiltonian system,
generated by the functional
$I^{\mu}$ (if it is not the momentum operator or annihilator of
bracket (\ref{bracket}), let us remind that all these systems have
the common family of m - phase solutions).

All these flows commute with each other because of the commutation of
functionals $\int J^{\mu}(X)dX$ with respect to Dubrovin -
Novikov bracket, and, besides that, the same is valid for the flows
generated by the integrals with respect to $X$ of the averaged
densities of all functionals, having the form (\ref{laws}) and
commuting with the Hamiltonian and integrals $I^{\nu}$,
since any of these functionals can be included into the set
$\{I^{\nu}\}$ instead of any of presenting there
integrals $I^{\nu}$, and, according to Theorem 2, this will not
change the bracket (\ref{sdn}). The integrals with respect to $X$
of the averaged density of momentum operator (\ref{momentum}) and the
annihilators of bracket (\ref{bracket}),
having the form (\ref{laws}), generate in the bracket (\ref{sdn})
the shift with respect to $X$ and zero flows respectively.
The integral with respect to $X$ of the averaged density
of the Hamiltonian (\ref{hamilt}) generates
Whitham's equations for m - phase solutions of system
(\ref{system}).

In closing, the author expresses his gratitude to S.P.Novikov,
who suggested the above problem, for
his attention to the work and also to V.L.Alexeev, O.I.Mokhov,
M.V.Pavlov and E.V.Ferapontov for fruitful discussions.

The research
was supported by Russian Foundation for Fundamental Research,
grant 96-01-01623, grant INTAS N 96-0770 
and the Landau Scholarship
awarded by KFA
Forschungszentrum J\"ulich GmbH.


\begin{thebibliography}{99}

\bibitem{with} G. Whitham. Linear and Nonlinear Waves,
Wiley, New York (1974).

\bibitem{luke} Luke J.C. A perturbation method for
nonlinear dispersive wave problems. Proc. Roy. Soc. London Ser. A,
{\bf 292}, No. 1430, 403-412 (1966).

\bibitem{dn1} B.A.Dubrovin and S.P.Novikov.
"Hamiltonian formalism of one-dimensional systems of hydrodynamic
type and the Bogolyubov - Whitham averaging method."
Dokl. Akad. Nauk SSSR, {\bf 270}, No. 4,
781-785 (1983).

\bibitem{dn2} B.A.Dubrovin and S.P.Novikov.
"Hydrodynamics of weakly deformed soliton lattices.
Differential geometry and Hamiltonian theory."
Uspekhi Mat.Nauk, {\bf 44}, No. 6 (270), 29-98 (1989).

\bibitem{dn3} B.A.Dubrovin and S.P.Novikov.
Hydrodynamics of soliton lattices.
Sov. Sci. Rev. C, Math. Phys. 1993, V.9. part 4.
P. 1-136.

\bibitem{dm} S.Yu. Dobrokhotov and V.P.Maslov.
Finite-Gap Almost Periodic Solutions in the WKB
Approximation. Contemporary Problems in Mathematics [in Russian],
Vol. 15, Itogi Nauki i Tekhniki, VINITI, Moscow (1980).

\bibitem{tsarev} S.P.Tsarev. "On Poisson brackets and
one-dimensional Hamiltonian systems of hydrodynamic type."
Dokl. Akad. Nauk SSSR, {\bf 282}, No. 3, 534-537 (1985).

\bibitem{fer1} E.V.Ferapontov. On integrability of $3 \times 3$
semi-Hamiltonian \linebreak hydrodynamic type systems
$u^{i}_{t} = v^{i}_{j}(u) u^{j}_{x}$ which do not possess Riemann
invariants. Physica D 63 (1993) 50-70 North-Holland.

\bibitem{fer2} E.V.Ferapontov. On the matrix Hopf equation and
integrable Hamiltonian systems of hydrodynamic type, which do not
possess Riemann invariants. Physics Letters A 179 (1993) 391-397
North-Holland.

\bibitem{novmal} S.P.Novikov and A.Ya.Maltsev.
"The Liouville form of averaged Poisson brackets."
Uspekhi Mat. Nauk, {\bf 48}, No. 1 (289), 155-156 (1993).

\bibitem{mokhov1} O.I.Mokhov.
Uspekhi Mat. Nauk. Vol.40, N 5, 257-258 (1985).

\bibitem{mokhov2} O.I.Mokhov.
Funk. Analiz i ego pril. Vol. 21, N 3, 53-60 (1987).

\bibitem{grinberg} N.I.Grinberg. 
Uspekhi Mat. Nauk. Vol.40, N 4, 217-218 (1985).

\bibitem{malpav} A.Ya.Maltsev and M.V.Pavlov.
"On Whitham's Averaging Method."
Functional Analysis and Its Applications, Vol. 29, No. 1, 7-24 (1995).

\end{thebibliography}
\end{document}